\documentclass[twocolumn]{svjour3}  

\usepackage{amsmath,amssymb,amsfonts}
\usepackage{graphicx}
\usepackage{textcomp} 
\usepackage[utf8]{inputenc}
\usepackage[T1]{fontenc}

\usepackage{ifthen}
\usepackage{hyperref}

\usepackage[dvipsnames,table]{xcolor}
\usepackage{minted}
\usepackage{soul}
\usepackage[normalem]{ulem} 
\usepackage{subcaption}
\usepackage{epsf,picinpar}
\usepackage{varioref}
\usepackage{algorithm}
\usepackage[noend]{algpseudocode} 
\usepackage{fdsymbol}
\usepackage{tcolorbox}
\usepackage{forest} 
\usepackage{enumerate}
\usepackage[framemethod=TikZ]{mdframed}
\usepackage{framed}

\usepackage{pifont}
\usepackage[pscoord]{eso-pic}

\usepackage[authoryear,round]{natbib}
\setcitestyle{authoryear,open={(},close={)},aysep={ },yysep={,},notesep={; }}
\usepackage{booktabs}
\usepackage{multirow}
\usepackage{float}
\usepackage{array}
 \usepackage{tabularx}
\usepackage{harveyballs}
\usepackage{makecell}
\usepackage{wrapfig}
\usepackage{svg}
\usepackage{orcidlink}
\usepackage{listings}
\usepackage{tikz}
\usepackage{calc}
\usepackage{siunitx}
\usepackage{hyperref}
\newcommand{\secref}[1]{Section~\ref{#1}}

\newcommand{\figref}[1]{Fig.~\ref{#1}}

\newcommand{\tabref}[1]{Table~\ref{#1}}

\newcommand{\todo}[1]{
    \ifthenelse{\boolean{showannotations}}%
    {\ifthenelse{\equal{#1}{}}{\textcolor{red}{TODO}}{\textcolor{red}{TODO:~{#1}}}}%
    {}%
}

\newcommand{\assignedto}[1]{%
    \ifthenelse{\boolean{showannotations}}%
    {\textbf{\noindent\ding{46}\textcolor{white}{~\colorbox{\assignementcolor}{Assigned to:}}~\textcolor{\assignementcolor}{#1}\\}%
    }
    {}
}

\newcommand{\rem}[1]{%
    \ifthenelse{\boolean{showannotations}}%
    {\textcolor{\oldtextcolor}{\st{#1}}}%
    {}%
}

\newcommand\add[1]{%
    \ifthenelse{\boolean{showannotations}}%
    {\textcolor{\newtextcolor}{{#1}}}%
    {#1}%
}

\newcommand\rep[2]{%
    \ifthenelse{\boolean{showannotations}}%
    {\rem{#1}~\add{#2}}%
    {#2}%
}

\newcommand{\darkmodecontrol}{%
    \ifthenelse{\boolean{darkmode}}%
    {\pagecolor{black!20}
    \color[rgb]{0,0,0} 
    }%
    {}%
}

\newcolumntype{L}[1]{>{\raggedright\let\newline\\\arraybackslash\hspace{0pt}}m{#1}}
\newcolumntype{C}[1]{>{\centering\let\newline\\\arraybackslash\hspace{0pt}}m{#1}}
\newcolumntype{R}[1]{>{\raggedleft\let\newline\\\arraybackslash\hspace{0pt}}m{#1}}

\floatstyle{ruled}
\newfloat{algo}{tbp}{loa}
\floatname{algo}{Algorithm}

\algnewcommand\algorithmicswitch{\textbf{switch}}
\algnewcommand\algorithmiccase{\textbf{case}}
\algnewcommand\algorithmicassert{\texttt{assert}}
\algnewcommand\Assert[1]{\State \algorithmicassert(#1)}%
\algdef{SE}[SWITCH]{Switch}{EndSwitch}[1]{\algorithmicswitch\ #1\ }{\algorithmicend\ \algorithmicswitch}%
\algdef{SE}[CASE]{Case}{EndCase}[1]{\algorithmiccase\ #1}{\algorithmicend\ \algorithmiccase}%
\algtext*{EndSwitch}%
\algtext*{EndCase}%
\algrenewcommand\algorithmicindent{1em}

\newtcolorbox{answerbox}{
  colback=gray!10,     
  colframe=black,      
  boxrule=0.3pt,       
  arc=3pt,             
  width=\columnwidth,
  boxsep=3pt,          
  left=3pt,            
  right=3pt,           
  top=3pt,             
  bottom=3pt           
}

\newboolean{showannotations}
\setboolean{showannotations}{true} 

\newboolean{shownames}
\setboolean{shownames}{true}

\newboolean{darkmode}
\setboolean{darkmode}{false} 

\newcommand{\newtextcolor}{blue}
\newcommand{\oldtextcolor}{red}
\newcommand{\assignementcolor}{orange}
\definecolor{highlightcolor}{rgb}{.99, 1, .0}
\sethlcolor{highlightcolor}
\definecolor{orcidlogocol}{HTML}{A6CE39}

\mdfsetup{%
   backgroundcolor=gray!3,
   roundcorner=3pt%
}

\newlength\WIDTHOFBAR
\lstdefinestyle{sqlstyle}{
    language=SQL,
    basicstyle=\ttfamily\small,
    keywordstyle=\color{blue}\bfseries,
    stringstyle=\color{orange},
    commentstyle=\color{gray}\itshape,
    morekeywords={MATCH,RETURN,UNWIND,CALL,YIELD,DETACH,WITH,OPTIONAL,to}, 
    showstringspaces=false,
    frame=single,
    columns=flexible,
    breaklines=true,
    xleftmargin=4mm,
    xrightmargin=4mm
}

\lstdefinelanguage{Kotlin}{
  morekeywords={package,import,class,interface,fun,object,val,var,typealias,
    is,in,Int,Double,Float,Boolean,String,Unit,Nothing,List,enum},
  sensitive=true,
  morecomment=[l]{//},
  morecomment=[s]{/*}{*/},
  morestring=[b]",
}

\definecolor{codegray}{gray}{0.35}
\definecolor{codegreen}{RGB}{40,120,40}
\definecolor{codeblue}{RGB}{40,70,150}
\lstdefinestyle{pythonstyle}{
    language=Python,
    basicstyle=\ttfamily\scriptsize,
    keywordstyle=\color{codeblue},
    commentstyle=\color{codegray},
    stringstyle=\color{codegreen},
    columns=fullflexible,
    keepspaces=true,
    showstringspaces=false,
    breaklines=false,
    aboveskip=0pt,
    belowskip=0pt
}

\begin{document}


\title{Does AI-Assisted Development Homogenize Software? An Empirical Study of Code Similarity, Quality, and Evolution}



\author{Luciano Marchezan         \and
        Romain Lefeuvre  \and
        Kevin Delcourt \and\\
        Benoit Combemale \and
        Houari Sahraoui
}


\institute{Luciano Marchezan \at
              Université de Montréal, Montréal, Canada\\
              University of York, York, UK \\
              \email{lucianomarchp@gmail.com}\\
              ORCID: \href{https://orcid.org/0000-0003-3096-580X}{0000-0003-3096-580X}
           \and
           Romain Lefeuvre \at
              Université de Rennes, Rennes, France\\
              \email{romain.lefeuvre@inria.fr}\\
              ORCID: \href{https://orcid.org/0009-0008-2019-1058}{0009-0008-2019-1058}
           \and
           Kevin Delcourt \at
              Université de Montréal, Montréal, Canada\\
              \email{kevin.delcourt@umontreal.ca}\\
              ORCID: \href{https://orcid.org/0009-0005-2988-7308}{0009-0005-2988-7308}
           \and
           Benoit Combemale \at
              Université de Rennes, Rennes, France\\
              \email{benoit.combemale@inria.fr}\\
              ORCID: \href{https://orcid.org/0000-0002-7104-7848}{0000-0002-7104-7848}
           \and
           Houari Sahraoui \at
              Université de Montréal, Montréal, Canada\\
              \email{sahraouh@iro.umontreal.ca}\\
              ORCID: \href{https://orcid.org/0009-0000-6980-5454}{0009-0000-6980-5454}
}

\date{Received: date / Accepted: date}

\maketitle

\begin{abstract}
AI-assisted software development is evolving from interactive LLM assistance toward agentic systems that autonomously perform software engineering tasks, yet the effects of this shift remain insufficiently understood. We present an empirical study of 50 long-lived Java and Python open-source projects across three eras: pre-LLM, potentially LLM-assisted, and detected agentic development. Using snapshots from Software Heritage, we examine code similarity, software quality, and software evolution. Similarity is measured semantically through embedding-based cosine similarity and syntactically through CodeBLEU*. Quality is assessed with SonarQube metrics covering maintainability, complexity, reliability, and security. Evolution is characterized by annualized growth rates of LOC, functions, classes, files, and statements, and by commit activity in the three months preceding each snapshot. Results show no consistent evidence of increased similarity as semantic and syntactic similarity decreased significantly in Java for both the LLM and agentic eras relative to the pre-LLM era, while Python showed no significant differences. Software quality remained largely stable, although the agentic era showed significantly lower code-smell and SQALE index densities. Structural growth was generally lower in the agentic era, particularly for files, whereas development activity increased moderately in Java, especially in commits and added lines. These findings challenge the assumption that LLM- and agent-assisted development necessarily homogenizes code, degrades quality, or accelerates growth.


\end{abstract}
\section{Introduction}\label{sec:introduction} 
The rapid progress of Large Language Models (LLM) has led to their widespread adoption in software development. According to the 2025 Stack Overflow Developer Survey, 84\% of developers already use, or plan to use, AI-based tools~\citep{stackoverflowStackOverflowx2019s}. Their role has evolved from inline code completion, where a 2022 study of GitHub Copilot reported about 25\% of suggestions accepted without modification~\citep{ziegler2022productivity}, to autonomous agents producing complete pull requests~\citep{khan2026adoption}.

Despite this adoption, developers remain cautious about AI-generated code. One concern is \textbf{software similarity}, which we define as the degree to which similar programming goals are realized through similar implementation choices across projects and over time. LLM outputs tend to be more homogeneous than human-generated ones~\citep{anderson2024homogenization}, and in creative domains LLM assistance often reduces the diversity of ideas~\citep{moon2025homogenizing}. Whether similar effects emerge in software ecosystems is unclear.

The implications of increased similarity are also multi-faceted. If LLMs preferentially recommend widely adopted languages, libraries, design patterns, or conventions~\citep{twist2025study}, they may improve standardization, consistency, and maintainability, and AI-assisted development has been associated with better adherence to coding standards~\citep{Xu2026CodeTransformed}. Software engineering, however, has historically evolved through continual exploration of alternative abstractions, architectures, and implementation strategies~\citep{Lehman1980SoftwareEvolution,Brooks1987NoSilverBullet}, so excessive homogenization could reduce the ecosystem's exploratory capacity and limit novel solutions.

Beyond code similarity, another emerging concern is the potential impact of AI-assisted development on \textbf{software quality}. 
The same Stack Overflow survey~\citep{stackoverflowStackOverflowx2019s} found that only a minority of developers fully trust AI-generated outputs, while many report that validating and debugging generated code remains a significant challenge. 
These concerns are supported by a growing body of empirical evidence highlighting limitations of LLM-generated code, including maintainability issues, hallucinated or inconsistent artifacts~\citep{Liu2024BeyondFC}, security vulnerabilities~\citep{Yetistiren2023EvaluatingTC}, reliance on deprecated APIs~\citep{Wang2024LLMsML}, overcorrections during code modification tasks~\citep{Jin2026AreLR}, and deficiencies in automatically generated tests~\citep{Tang2023ChatGPTVS}. Likewise, studies comparing human-written and AI-generated code, or investigating AI-assisted development in real-world repositories, report mixed trade-offs between short-term productivity gains and longer-term quality indicators such as code complexity, duplication, warnings, and ownership concentration~\citep{He2026SpeedCostQuality,Agarwal2026AIIDEsAutonomousAgents,Tomaz2026ImpactsOG}.

In addition to these concerns, the increasing integration of LLMs raises questions about their implications for \textbf{software evolution}. Software systems continuously grow and change as requirements evolve, new capabilities are introduced, and existing functionality is maintained or replaced~\citep{Lehman2006,Assuncao2025}. As AI tools increasingly support tasks ranging from code generation to autonomous modification of software artifacts~\citep{Sauvola2024,Jin2024FromLT}, they may influence not only how individual functions are implemented but also the pace and structural trajectory of software development~\citep{Lehman1980SoftwareEvolution}. Recent empirical studies of agentic contributions document differences in commit activity, code additions and deletions, and files modified, as well as the types of structural changes performed during agentic refactoring~\citep{Ogenrwot2026,Horikawa2025}. 

Increased development activity, however, does not necessarily translate into sustained structural growth~\citep{Hatton2017}, nor does faster code production necessarily imply more extensive or frequent changes to existing systems. AI-assisted development may facilitate incremental modifications and refactoring, but it may also introduce additional code or prompt subsequent revisions (which may also affect similarity and quality). Hence, understanding these dynamics is particularly important for long-lived open-source projects, where the consequences of AI adoption unfold within established codebases shaped by years of development and maintenance.

These observations motivate the central question investigated in this paper: \textbf{whether the widespread adoption of LLM-assisted software development is associated with measurable changes in the similarity, quality, and growth of open-source software systems}. For that, we define three research questions (RQ): 

\begin{description}
    \item \textbf{RQ1.}To what extent does the code similarity of long-lived open-source software change across eras?
    \item \textbf{RQ2.} To what extent does the quality of long-lived open-source software change across eras considering different quality dimensions?
    \item \textbf{RQ3.} To what extent do the development and structural characteristics of long-lived open-source software change across eras?
\end{description}

To investigate these RQs, we conduct an empirical study of 50 long-lived, engineered open-source projects, comprising 25 Java and 25 Python repositories, retrieved from Software Heritage~\citep{Abramatic2018UniversalArchive}. We compare snapshots from 2021, 2024, and 2026, representing pre-LLM development, potentially LLM-assisted development, and development with detected traces of coding-agent adoption, respectively. The projects are selected from a dataset of repositories exhibiting evidence of agentic usage~\citep{Robbes2026AgenticMA}. For each snapshot, we examine code similarity through semantic and syntactic measures of function resemblance, and software quality through structural and static-analysis indicators of maintainability, complexity, reliability, and security. Finally, we investigate software evolution by measuring the annualized growth of source lines of code, functions, classes, files, and statements, as well as development activity through commit activity during the three months preceding each snapshot. 

The results show no consistent evidence of increased code similarity or substantial changes in software quality across the three development eras. Semantic and syntactic similarity decreased significantly ($p<0.05$) in Java for both the LLM and Agentic eras relative to the Pre-LLM era, while Python exhibited no statistically significant differences. Software quality remained largely stable, with the Agentic era showing statistically significant reductions ($p<0.05$) in code-smell and SQALE index densities. Structural growth was generally lower in the LLM-to-Agentic transition, particularly for files. Development activity showed a moderate increase in Java during the Agentic era, especially in commits and lines added, whereas Python exhibited little overall change.

Our findings challenge the assumption that LLM- and agent-assisted development necessarily homogenizes code, degrades software quality, or accelerates structural evolution and growth. Instead, the observed changes vary across programming languages and metrics, with encouraging improvements in selected maintainability indicators but no consistent changes in complexity, reliability, security, or structural growth. 

The main contributions of this paper are:  
\begin{enumerate} 
    \item A \textbf{curated dataset of 50 long-lived open-source projects}, comprising 25 Java and 25 Python repositories, with snapshots spanning pre-LLM, LLM-assisted, and agentic development eras;
    \item An \textbf{empirical characterization of code similarity} across pre-LLM, LLM-assisted, and agentic development eras, using complementary semantic and syntactic measures across Java and Python projects;
    \item A \textbf{longitudinal analysis of software quality} across maintainability, complexity, reliability, and security dimensions, identifying changes in static-analysis indicators across development eras;
    \item An \textbf{empirical investigation of software evolution}, examining structural growth and development activity to assess how these characteristics vary across development eras;
    \item \textbf{Evidence-based insights} into the relationship between LLM- and agent-assisted development and software evolution, highlighting language-dependent similarity trends, improvements in selected maintainability indicators, and the absence of consistent changes in structural growth or development activity;
\end{enumerate}

The rest of the paper is organized as follows. \secref{sec:background} presents the background and related concepts. \secref{sec:method} describes the research methodology, including project selection, data collection, and analysis procedures. \secref{sec:results} presents the empirical results, followed by their interpretation and implications in \secref{sec:discussion} and threats to validity in \secref{sec:threats}. Related work is described in \secref{sec:related}. Finally, \secref{sec:conclusion} concludes the paper and outlines directions for future work.
\section{Background}\label{sec:background}

\subsection{Evolution of LLM usage in Software Development}\label{sec:llmevolution}

\begin{figure}
  \includegraphics[width=\linewidth]{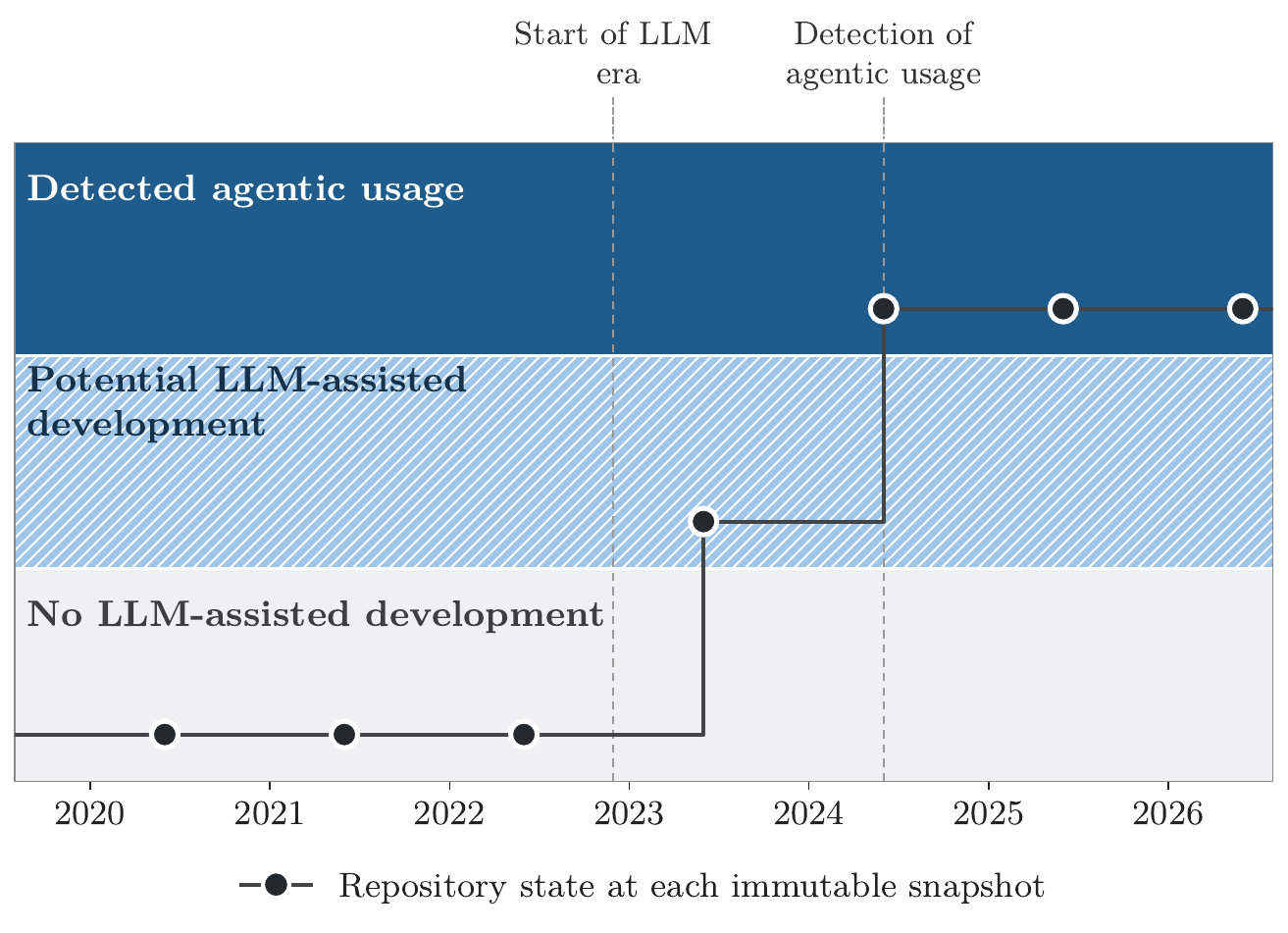}
  \caption{Example LLM-assisted development state evolution across a repository history }
\label{fig:exemple}
\end{figure}

The integration of LLMs into software development has evolved from interactive assistance and code completion to agents capable of executing multi-step development tasks~\citep{Jin2024FromLT}. These forms of assistance differ in their level of autonomy and in the evidence they leave in repository histories~\citep{Pan2024IDU,robbes2026promises}. To study software evolution across these changes, we distinguish three development eras, illustrated in \figref{fig:exemple}.

\paragraph{Pre-LLM development.}
This era covers repository snapshots from before the public release of ChatGPT in November 2022~\citep{OpenAI2022ChatGPT}. It provides a historical baseline, although it does not imply the complete absence of AI-assisted development, since tools such as GitHub Copilot were already available in technical preview from June 2021~\citep{GitHub2021Copilot}.

\paragraph{Potentially LLM-assisted development.}
This era begins with the public release of ChatGPT and extends until coding-agent activity is detected in a repository. LLM assistance during this period may include code completion, API explanations, design suggestions, and code generation followed by developer modification~\citep{Pan2024IDU}. Because such assistance is not necessarily recorded in version-control metadata, repository history alone cannot establish whether or to what extent individual contributions were LLM-assisted~\citep{robbes2026promises}. We therefore use potentially LLM-assisted as a temporal classification rather than as confirmation of AI use.

\paragraph{Detected agentic development.}
Coding agents can perform sequences of development actions, including inspecting repositories, editing files, executing tests, and preparing changes for review~\citep{Liu2026}. Because of this autonomy, their activity may leave observable traces, such as agent-attributed commits, pull requests, or collaboration metadata. Datasets and detection heuristics have been developed to identify such traces in public repositories~\citep{Li2026AIDev,robbes2026promises}. We classify a repository snapshot as belonging to this era when repository-specific evidence of coding-agent activity has been detected. This classification establishes observed agentic activity, but does not imply that every contribution in the snapshot was produced by an agent.

\subsection{Code Similarity}

We use the term \emph{code similarity} to denote the degree to which code fragments resemble one another in their representations or implemented functionality. This includes variation in algorithms, library and API choices, control-flow structures, decomposition strategies, naming conventions, idioms, comments, and tests~\citep{chon2024functional}. 

In this study, we focus on similarity within software projects. In particular, we distinguish (i) syntactic similarity, which concerns program structure; and (ii) semantic similarity, which concerns functionality. 
As such, two functions can be similar or different along any of those dimensions, and this underpins most code clone studies~\citep{Roy2009ComparisonCloneDetection}.

Neither similarity nor repetition has an intrinsically positive or negative quality interpretation. \citet{Kapser2008Cloning} document cloning practices that can support legitimate engineering decisions; and conversely, they document that greater variation does not necessarily imply a broader repertoire of useful solutions. Similarity is therefore a descriptive property of the codebase, which is distinct from a quality metric. This distinction motivates examining similarity and quality separately.

\subsection{Code Representations and Comparison Measures}\label{sec:code-representations}

Code comparison relies on representations that preserve different information. Syntactic representations such as token sequences and abstract syntax trees (ASTs) expose lexical and syntactic information~\citep{Roy2009ComparisonCloneDetection}, while learned representations instead encode source fragments as dense vectors whose geometry reflects patterns acquired during training.

In this context, code embedding models can be used to represent functions in a vector space and estimate their semantic similarity~\citep{Ben2018}. In our study, we use two models, CodeBERT and CodeT5. CodeBERT is a Transformer-based encoder pretrained on programming and natural languages, including Java and Python, and supports downstream code-related tasks~\citep{Feng2020CodeBERT}. CodeT5 uses an encoder--decoder architecture and code-specific pretraining objectives to support code understanding and generation~\citep{Wang2021CodeT5}. Instead of using their pre-trained variants, however, we adopt their versions fine-tuned for Type-IV code clone detection, which targets functions that implement similar functionality despite differences in their syntax and structure~\citep{Marchezan2026Kamino}.

Considering syntactic similarity, \emph{CodeBLEU} was introduced to evaluate generated code against reference implementations~\citep{CodeBLEU}. It combines lexical matching based on token $n$-grams, keyword-weighted matching, AST subtree matching, and data-flow matching. Although these components capture different aspects of code resemblance, the data-flow component incorporates a restricted semantic signal based on dependencies between variables. Therefore, CodeBLEU is not a purely syntactic measure. Alternatively, \emph{CodeBLEU*}~\citep{Marchezan2026Kamino} is computed by disabling the data-flow matching component, thereby measuring syntactic similarity by averaging the lexical, keyword-weighted, and AST subtree matching components of CodeBLEU. For this reason, we adopt CodeBLEU*.

\subsection{Indicators of Code Quality}

Code quality includes many dimensions such as functional correctness, security, maintainability, or complexity~\citep{NunesVarela2017}.
The empirical software engineering literature has developed many proxies for these dimensions, including code smells \citep{Zakeri-Nasrabadi2023ASL,Jabrayilzade2024TaxonomyOI}, 
or technical-debt models~\citep{Perera2024ASM,Halepmollasi2025PredictingVC}. 

Static-analysis tools such as the ones provided by SonarQube\footnote{SonarQube: \url{https://www.sonarsource.com/}} apply language-specific rules and report issues associated with reliability, security, and maintainability~\citep{SonarSource2026CloudDocumentation}. 
Their measurements include cyclomatic and cognitive complexity, issue counts,  estimated remediation effort, among many other metrics that are either language-specific or agnostic~\citep{SonarSource2026MetricDefinitions}. These indicators capture different properties related to one or multiple dimensions of code quality.

\subsection{Software Evolution}

Software evolution refers to the ongoing changes that software systems undergo as requirements evolve, new capabilities are introduced, and existing functionality is maintained or replaced~\citep{Lehman2006}. These changes can affect both the structure of a software system and the activities involved in its development. As LLMs increasingly support code generation, modification, and autonomous development tasks~\citep{Sauvola2024,Jin2024FromLT}, they may influence how software systems grow and evolve over time~\citep{Assuncao2025}. 
Structural growth describes changes in the size and organization of a software system, which can be characterized using measures such as lines of code, functions, classes, files, and statements~\citep{Henry1981}. 

Structural growth alone, however, does not capture the volume of development work performed during a period. Development activity instead characterizes the frequency and extent of changes made to a codebase. Measures such as commit frequency and lines added or removed provide complementary perspectives on this activity~\citep{Hatton2017}. 

\subsection{Historical Source-Code Observations}

Historical studies require identifiable source versions and sufficient provenance to recover the same artifacts. 
Software Heritage (SWH) addresses this need through an independent, long-term archive of publicly available source code and its development history~\citep{Abramatic2018UniversalArchive}. It aggregates artifacts from multiple hosting platforms and package repositories in a common representation~\citep{Pietri2020SoftwareHeritageGraph}.
To date, SWH is the largest public archive of Open Source Software source code and development history, with more than 443 million repositories.

The SWH data model is a deduplicated Merkle directed acyclic graph (DAG) whose nodes represent source code artifacts, including commits and releases. Each node is identified by a unique and persistent \emph{SoftWare Hash IDentifier} (SWHID).\footnote{\url{https://swhid.org}, standardized as ISO/IEC 18670:2025.} Repositories are represented as \emph{origins}, each time an origin is archived, SWH records a \emph{snapshot} that references the commits at the heads of its branches. Unlike a Git repository, where branches can be moved and history rewritten, the SWH graph is immutable and resistant to history alteration. Re-archiving a repository adds a new snapshot and leaves earlier snapshots, and the commits they reference, untouched.

Relative to collecting repositories directly from GitHub, SWH's principal advantages are preservation, model immutability, cross-platform normalization, and persistent identification. This makes it particularly suitable as the source-artifact layer of a reproducible historical study~\citep{lefeuvre2023fingerprinting}, while forge data can remain complementary to study development activity.

\subsection{Motivation and Research Gap}

Although existing studies have investigated the characteristics of LLM-generated code and the effects of AI-assisted development, their findings span different settings, tasks, and quality indicators, making the implications for long-lived software systems difficult to establish~\citep{He2026SpeedCostQuality,Tomaz2026ImpactsOG}. In particular, findings about code generated in isolated tasks do not necessarily generalize to established repositories, where new contributions interact with existing implementations and accumulated design decisions~\citep{Agarwal2026AIIDEsAutonomousAgents}. Moreover, changes in similarity, quality, and development activity may be related but do not necessarily follow the same direction. A longitudinal analysis combining these perspectives is therefore needed to characterize how these properties vary across the transition from pre-LLM development to LLM-assisted and agentic development. Next, we present the methodology to address this gap.

\section{Research Methodology}\label{sec:method}

The goal of this study is to investigate whether the \textbf{widespread adoption of LLM-assisted}, and more recently of \textbf{agentic} software development, is associated with \textbf{measurable changes in the similarity, quality, and growth of open-source software systems}.
Across the three eras of software development discussed in \secref{sec:llmevolution}, i.e., \textit{pre-LLM development}, \textit{potentially LLM-assisted development}, and \textit{detected agentic development}, we formulate the following research questions:

\begin{description}
\item[\textbf{RQ1 - Code Similarity:}] To what extent does the code similarity of long-lived open-source software change across eras?
\begin{description}
\item[\textbf{RQ1.1} - Semantic similarity:] We measure the cosine similarity of function pairs.
\item[\textbf{RQ1.2} - Syntactic similarity:] We measure CodeBLEU* of function pairs.
\end{description}

\item[\textbf{RQ2 - Software Quality:}] To what extent does the quality of long-lived open-source software change across eras considering different quality dimensions?
\begin{description}
    \item[\textbf{RQ2.1} - Maintainability:] Code Smells, Duplicated Lines, SQALE Index, and Comment Lines. 
    \item[\textbf{RQ2.2} - Complexity:] Cyclomatic and Cognitive Complexity.
    \item[\textbf{RQ2.3} - Reliability:] Bugs.
    \item[\textbf{RQ2.4} - Security:] Vulnerabilities and Security Hotspots. 
\end{description}

\item[\textbf{RQ3 - Software Evolution:}] To what extent do the development and structural characteristics of long-lived open source software change across eras?

\begin{description}
\item[\textbf{RQ3.1 - Structural Growth:}] We measure the structural growth by the annualized growth of NCLOC, functions, classes, files, and statements.
\item[\textbf{RQ3.2 - Development Activity:}] We measure the development activity by the number of commits and the number of added and removed lines of code in the three months preceding each snapshot era.
\end{description}

\end{description}

Details on rationale, method, and metrics for each RQ are described in \secref{sec:analysis}.

\subsection{Dataset Curation}

To answer our RQs, we target a population of \textit{engineered software projects} that transitioned to agentic development over their history. Specifically, we focus on long-lived open-source projects written in Java and Python. For this, we adopt the dataset provided by \cite{Robbes2026AgenticMA}, composed of 27,192 GitHub repositories that exhibit traces of agentic development.
This dataset covers 63 coding agents and relies on 196 heuristics that leverage different kinds of traces in configuration files, commit authorship, or interactions in collaborative development workflows in order to detect agentic activities.
We then leverage the SWH archive to apply our selection criteria, retaining only long-lived \textit{Engineered Software Projects}, and to extract an immutable snapshot of each repository for each era previously introduced.
Finally, we performed stratified random sampling by programming language on the filtered repositories to obtain a final sample of 50 repositories.

\paragraph{Sampling strategies.}

We follow a two-phase sampling strategy, using as our initial sampling frame the repositories from the dataset introduced by \citet{Robbes2026AgenticMA}. Specifically, the initial sampling frame consists of the repositories from that dataset that had been archived by SWH and were available in the 2026-06-04 SWH export, representing 13,123 of the 27,192 repositories in the original dataset. This gap is due to SWH’s ingestion delay, which can take several months before newly created repositories are listed and archived.
Since this study explicitly targets long-lived repositories, it is not impacted by this limitation, as the missing repositories are expected to correspond to recent projects that would not have survived our sampling strategy.
To confirm this intuition, we randomly sampled 1,000 repositories absent from the SWH export and retrieved their creation date via the GitHub API. Of the 983 repositories still available on GitHub, 87.3\% were created after 2021-01-01 and would therefore have been filtered out by our sampling strategy.

We design the sampling workflow using the \textit{SamplingWorkflowDSL} introduced by \citet{lefeuvre2026modeling}, which supports reproducible and explicit description of the sample extraction.
The sampling workflow and code used to extract metadata from SWH are available in the replication package~\citep{EMSEGenAIReplication}.

\figref{fig:sampling_workflow} illustrates the sampling and selection process, reporting the number of filtered repositories at each stage.
We first select \textit{long-lived engineered software projects} by complementing the criteria of the original dataset with criteria commonly used in the literature. Beyond the 10-star threshold inherited from that dataset, we rely on intrinsic properties of repositories, as \cite{maj2024fault} identified limitations in relying on popularity metrics such as stars alone.
We apply the following criteria, all computed from the 2026-06-04 SWH export:

\begin{figure*}[th]
\centering
\begin{subfigure}[b]{\textwidth}
\centering
\includegraphics[width=\textwidth]{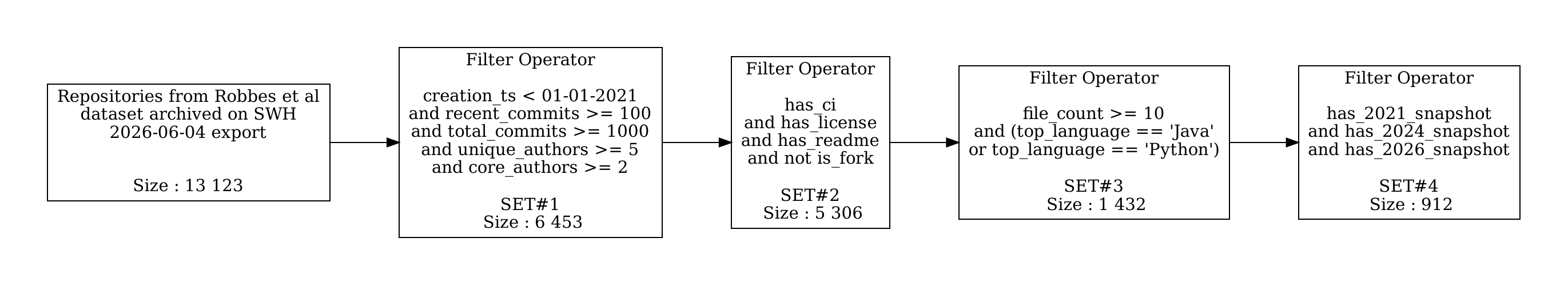}
\caption{Filtering \textit{long-lived engineered software projects} using Java and Python language}
\label{fig:workflow_filter}
\end{subfigure}
\begin{subfigure}[b]{\textwidth}
\centering
\includegraphics[width=\textwidth]{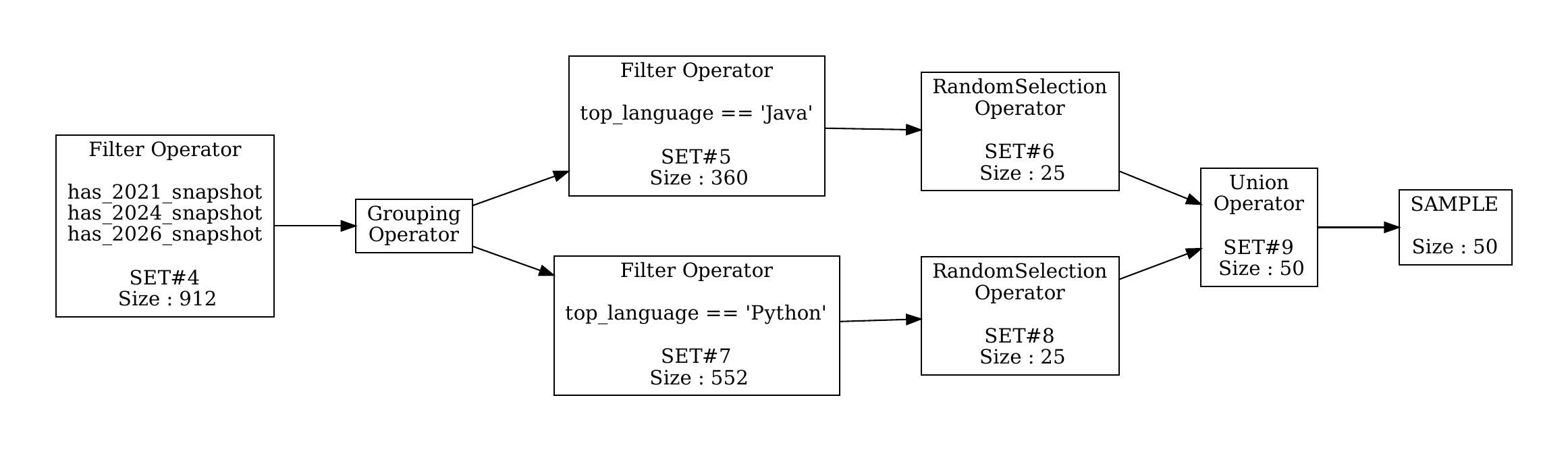}
\caption{Extracting a stratified random sample}
\label{fig:workflow_sampling}
\end{subfigure}
\caption{Sampling workflow used in this study, reporting the number of repositories filtered at each stage.}    \label{fig:sampling_workflow}
\end{figure*}

\textbf{Activity.} We filter repositories created before 2021 that have at least 1,000 commits (at least 10 of them since 2022-11-01, the start of the LLM era), at least five distinct commit authors, and at least two core authors, where core authors form the smallest group of authors whose commits together account for 80\% of all commits~\citep{munaiah2017curating}. \emph{Rationale.} Studying the impact of coding-agent adoption on code similarity and software quality requires projects with a long development history. The creation-date constraint ensures that every project is at least five years old, and the commits since 2021 ensure that it remained active in the LLM era. Requiring two core authors also excludes projects in which a single developer authored most of the commits.

 \textbf{Engineering practices.} We filter repositories with a license, a continuous integration and delivery (CI/CD) descriptor file~\citep{munaiah2017curating}, and a \texttt{README} file. We also exclude shared-commit forks~\citep{pietri2020forking} that the GitHub fork relation used to build the original dataset does not capture. \emph{Rationale.} These constraints provide insight into DevOps practices (CI/CD descriptor), documentation (\texttt{README}), and reuse conditions (license). Excluding forks also prevents us from analyzing the same development history several times, as forks share part of their history with another repository.

\textbf{Codebase.} We keep projects whose main language, determined from file extensions, is Python or Java, and that contain at least 10 files.  \emph{Rationale.} The file threshold excludes trivial codebases.

\textbf{Snapshot availability.} We keep projects with at least one SWH snapshot in each of 2021 (\textit{Pre-LLM}), 2024 (\textit{LLM}), and 2026 (\textit{Agentic}). For each year, we use the latest available snapshot.  

\figref{fig:sampling_workflow} reports the number of repositories remaining after each stage. These filters yield 912 long-lived Java and Python projects. \figref{fig:workflow_sampling} describes the second phase of our sampling strategy, where we apply stratified random sampling with strata divided by the main language of each repository, Java or Python.

\paragraph{Sample Representativeness.} To assess the representativeness of our sample, we check that it is both \textit{large and random}, and that it is of a \textit{similar distribution} to its stratum. Regarding the first argument, the size of the sample drawn from each stratum is determined using Cochran’s formula \citep{cochran1977sampling}.
We set a 95\% confidence level and a 20\% margin of error, assuming maximum variability ($p = 0.5$).
This margin is wider than the 5\% commonly adopted in empirical software engineering, but it allows for a smaller resulting sample, keeping the cost of the subsequent analyses manageable.
Under these parameters, the minimum sizes are 23 for the Java stratum and 24 for the Python stratum, rounded up. We draw 25 repositories from each stratum, which yields a final sample of 50 repositories.

Regarding the second argument, we compared the distributions of metadata of interest before and after random sampling. We used three metadata: the number of commits, the number of distinct commit authors, and the number of files. The comparison relies on two-sample Kolmogorov-Smirnov tests \citep{massey1951kolmogorov}, which are sensitive to differences in the shape of distributions. None of the six tests detects a statistically significant difference at $\alpha = 0.05$. The $p$-values for the Java stratum are 0.91 for commits, 0.85 for authors, and 0.98 for files, and those for the Python stratum are 0.24, 0.79, and 0.42, respectively.
Together, these two arguments support the representativeness of each sample with respect to its stratum.

\begin{table*}[t]
\centering
\addtolength{\tabcolsep}{-3pt}
\caption{Descriptive statistics of the sample (N=50). Statistics related to project history are computed on the default branch. }
\label{tab:sample-description}
\begin{tabular}{lcrrrrrrr}
\toprule
\textbf{Metric} & \textbf{Distribution} & \textbf{Min} & \textbf{Q1} & \textbf{Median} & \textbf{Mean} & \textbf{Q3} & \textbf{Max} & \textbf{StD.} \\
\midrule
Creation date & \includegraphics[width=1.4cm,height=0.28cm]{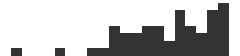} & 2000-12-01 & 2012-03-16 & 2015-04-12 & 2014-07-04 & 2018-01-06 & 2019-12-06 & 1480 days \\
Total commits & \includegraphics[width=1.4cm,height=0.28cm]{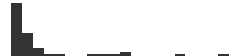} & 1016 & 1916.8 & 3990.5 & 9181.2 & 7517.5 & 65931 & 13551.6 \\
Unique authors & \includegraphics[width=1.4cm,height=0.28cm]{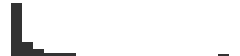} & 17 & 45.5 & 109.5 & 213.0 & 222.5 & 2394 & 361.5 \\
Core authors & \includegraphics[width=1.4cm,height=0.28cm]{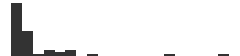} & 2 & 4.2 & 9.5 & 19.7 & 17.5 & 169 & 30.2 \\
Total files count & \includegraphics[width=1.4cm,height=0.28cm]{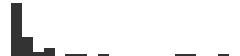} & 85 & 482.2 & 968 & 2685.6 & 2046.2 & 22158 & 4686.5 \\
Estimated size (MB) & \includegraphics[width=1.4cm,height=0.28cm]{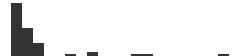} & 0.5 & 6.4 & 27.1 & 58.5 & 52.2 & 543.1 & 100.7 \\
\midrule
Code files count & \includegraphics[width=1.4cm,height=0.28cm]{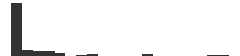} & 24 & 175.8 & 517 & 1574.8 & 1433.8 & 17054 & 2926 \\
NCLOC & \includegraphics[width=1.4cm,height=0.28cm]{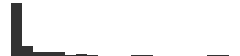} & 2592 & 22823 & 61742.5 & 198195.4 & 180485 & 2183986 & 362779.7 \\
Classes & \includegraphics[width=1.4cm,height=0.28cm]{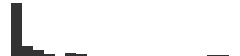} & 23 & 195.2 & 601.5 & 1795.4 & 1739.2 & 25614 & 3787.5 \\
Functions & \includegraphics[width=1.4cm,height=0.28cm]{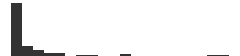} & 135 & 1295.2 & 3832.5 & 12368.1 & 11970.5 & 148943 & 23779.8 \\
Statements & \includegraphics[width=1.4cm,height=0.28cm]{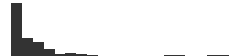} & 2007 & 10706.2 & 33002 & 70812.4 & 83634 & 646401 & 112803.8 \\
\bottomrule
\end{tabular}
\end{table*}
\tabref{tab:sample-description} reports descriptive statistics for the 50 repositories of our sample. The projects are long-lived, with creation dates ranging from 2000 to 2019 and a median creation date of 2015-04-12.
They also exhibit a substantial development history, with a minimum of 1,016 commits and a median of 3,990.5, and codebases that range from 85 to 22,158 files, or from 0.5 to 543.1 MB, with a median of 968 files for 27.1 MB. The sample also exhibits strong collaborative development. Every repository has at least 17 distinct commit authors, with a median of 109.5, and at least two core authors, with a median of 9.5.

Furthermore, the number of source code files (bottom part of \tabref{tab:sample-description}), ranged from 24 to 17054 (mean of 1574.8), with the Non-Comment Lines of Code (NLOC), number of classes, functions, and statements following a similar pattern.
All these statistics confirm that our sample is composed of \textit{long-lived engineered software projects}.

\subsection{Analysis Procedure}\label{sec:analysis}

In this section, we describe the metrics, techniques, and tools adopted to analyze and report the data and results to answer our RQs.

\subsubsection{RQ1. Code Similarity Analysis}
\label{sec:diversity-analysis}
To investigate RQ1, we analyze both semantic (RQ1.1) and syntactic similarity (RQ1.2) at the function level.  First, we extract the functions and create the corresponding pairs for both RQs.

\paragraph{Function extraction.}

For each repository snapshot, all functions are extracted using language-specific parsers.
Before the similarity analysis, a \textit{preprocessing} step is applied to remove functions that are unlikely to provide meaningful information. Specifically, common boilerplate methods, including getters, setters, and entry-point methods (e.g., \texttt{main}), are excluded using lightweight heuristics based on function names and structural characteristics.
A function is excluded if it satisfies all of the following criteria: (i) it contains fewer than two NLOC, excluding its signature and comments; (ii) it contains fewer than five tokens; and (iii) its name is ``main'' or starts with ``get’’ or ``set’'. Although these filters may exclude some potentially relevant functions, a substantial number of function pairs were retained for each project, as mentioned next.

After preprocessing, we isolate the functions that existed in the first snapshot (pre-LLM) from those introduced in the second (LLM era) and third (Agentic era) snapshots. To do so, we apply a simple \textit{diff filter} that retains only functions newly added between each era. The filtering is based on the function signature. For Java, the class name, method name, and parameter names and types are considered. For Python, the file name, function name, and number of parameters are considered. This difference is due to the limited capabilities of the Python parser. As with the preprocessing, this diff filter may exclude potential cases and consequently reduce the number of functions considered. For the LLM and Agentic eras, the filtering reduced the number of functions $\approx$ 80\% overall. In total, the number of functions remaining after this filter was $\approx$1M for the pre-LLM (avg. of 20K per project), $\approx$100K for LLM era (avg. of 2K per project), and $\approx$90K for the Agentic era (avg. of 1.8K per project). These represent a relatively large sample considering that the analysis is based on function pairs, as described next.


\paragraph{Pair selection.}
Exhaustively creating all possible function pairs would result in quadratic complexity, making large-scale pair-wise comparisons computationally infeasible.

To address this challenge, we adopt a pairing strategy inspired by large-scale code search and clone detection approaches~\citep{Chochlov2022}.
Each function is embedded using a pretrained CodeBERT model. These embeddings are used for candidate retrieval and are indexed using an Approximate Nearest Neighbor (ANN) structure. For each function, the top-(k) ANNs are retrieved based on cosine similarity in the embedding space. This process generates a reduced set of candidate pairs focusing on those with potentially high similarity while avoiding the need to enumerate all possible function combinations.

The value of (k) is selected empirically through a validation study. For that, we analyzed a subset of the repositories using exhaustive pairwise comparison to establish a reference set of close pairs. We then evaluated the candidates retrieval using different values of (k) (i.e., 10, 20, 50, 100), and retrieval effectiveness is measured using Recall@\emph{k}.
The selected value (i.e., 20) corresponds to the smallest (k) achieving a target recall (e.g., at least (95\%)), thereby minimizing computational cost while preserving the vast majority of similar pairs. After selecting the functions and creating the pairs, we obtained a total of $\approx$5.3M function pairs for pre-LLM, $\approx$1.3M for the LLM era, and $\approx$1.3M for the Agentic era, averaging between $\approx$13K and $\approx$200K function pairs per project.

\paragraph{RQ1.1. Semantic similarity analysis.}
Specifically for RQ1.1, we perform the semantic similarity analysis by comparing the function pair embeddings. For that, we used two versions of CodeBERT and CodeT5, which were finetuned for Type-IV clone detection (i.e., to capture semantic similarity)~\citep{Marchezan2026Kamino}. This distinction is relevant as these finetuned versions are more adequate to capture semantic representations when compared to their pre-trained versions (evidenced in an empirical evaluation~\citep{Marchezan2026Kamino}). For each pair, given two functions, $f_1$ and $f_2$, with embeddings $z_1$ and $z_2$, semantic similarity is computed using cosine similarity:

\begin{equation}
\text{sim}(z_1,z_2)=
\frac{z_1 \cdot z_2}
{|z_1||z_2|}.
\end{equation}

After the calculation, we record the similarity scores produced by both models and use them as complementary measures of semantic similarity.
The rationale is to obtain two perspectives on the similarity between function pairs, as the two models may capture different aspects of the semantic relationship between functions.

\paragraph{RQ1.2. Syntactic similarity analysis.}
For RQ1.2, we perform an analysis to characterize the degree of syntactic similarity between each function pair. For this purpose, as discussed in \secref{sec:code-representations}, we adopt CodeBLEU*~\citep{Marchezan2026Kamino}, which excludes the data-flow component to measure syntactic similarity. CodeBLEU* produces a similarity score in the range $[0,1]$, where higher values indicate greater syntactic resemblance.

\paragraph{Statistical Testing.}

We use the same statistical testing procedure for both RQ1.1 and RQ1.2. The descriptive analysis considers all function pairs to characterize the distribution of individual similarity scores without aggregation. For statistical testing, we aggregate pair-level scores using the median for each project, development era, programming language, and similarity measure. This approach ensures that each project contributes equally to the statistical tests, regardless of the number of function pairs it contains, while reducing the influence of extreme pair-level similarity scores that could disproportionately affect the mean.

We then apply paired Wilcoxon signed-rank tests to compare the three development eras (LLM versus Pre-LLM, Agentic versus Pre-LLM, and Agentic versus LLM), retaining only projects with available measurements in all three eras for the corresponding language and similarity measure. Effect sizes are measured using the paired rank-biserial correlation, which captures the direction and magnitude of the ranked paired differences. We interpret effect magnitudes with the rank-biserial correlation as negligible ($|r_{\mathrm{rb}}|<0.10$), small ($0.10\leq|r_{\mathrm{rb}}|<0.30$), moderate ($0.30\leq|r_{\mathrm{rb}}|<0.50$), or large ($|r_{\mathrm{rb}}|\geq0.50$). 
To account for multiple comparisons, $p$-values are adjusted using the Benjamini–Hochberg procedure, with a significance level of $\alpha=0.05$. The descriptive distributions of pair-level similarity scores and the project-level inferential results are reported separately, as they summarize different units of analysis.

\subsubsection{RQ2. Code Quality Analysis}
\label{sec:quality-metrics}
To assess software quality (RQ2), we adopt a practitioner-oriented measurement strategy based on metrics provided by \textit{SonarQube Community Edition}~\citep{SonarSource2026CloudDocumentation}.
We use its metrics to characterize quality properties that are directly observable during software development~\citep{Yu2023}.

\paragraph{Quality Metrics.}
The quality-related metrics adopted are: \emph{Cyclomatic Complexity}, \emph{Cognitive Complexity}, \emph{Bugs}, \emph{Vulnerabilities}, \emph{Security Hotspots}, \emph{Code Smells}, \emph{Duplicated Lines Density (DLD)}, \emph{SQALE Index}, and \emph{Comment Lines Density (CLD)}. A detailed description of each metric is available in our replication package~\citep{EMSEGenAIReplication}.

Because the absolute values of the majority of these metrics are naturally affected by repository size, directly comparing their raw values across projects may confound changes in software quality with differences in the amount of code being analyzed, rather than differences associated with the development era. To reduce this size-related bias, we normalize each metric $m$ by the total number of functions in the corresponding project:
\begin{equation}
m_{\mathrm{density}} =
\frac{m}{\#Functions}.
\end{equation}

This normalization provides a density measure that facilitates comparisons across projects of different sizes. 
The only exceptions are DLD and CLD, which are already calculated by SonarQube as density-based measures. We therefore keep their original values rather than applying our normalization. Since DLD and CLD are computed using a different normalization logic, their absolute values are not directly comparable with those of the other density metrics. This difference, however, does not affect our RQs, as our analysis focuses on comparing the same metric across development eras rather than comparing values across different metrics. 

\paragraph{Comparison procedure.}
For each repository, we compare the three snapshots (pre-LLM, LLM, Agentic). This design reduces the influence of repository-specific characteristics, such as domain, architecture, and coding conventions.
To simplify interpretation, we adopt a common reporting convention for quality-related metrics: positive changes indicate a greater presence of quality issues (i.e., decreased quality), whereas negative changes indicate a reduction in quality issues (i.e., improved quality).
Lastly, we organize these measurements from a higher-level perspective, by following the quality dimensions identified by \citet{NunesVarela2017}. Code Smells, DLD, SQALE Index, and CLD provide indicators related to \emph{Maintainability}~(RQ2.1); Cyclomatic Complexity and Cognitive Complexity characterize the \emph{Complexity} dimension~(RQ2.2); Bugs capture \emph{Reliability}~(RQ2.3); and Vulnerabilities, and Security Hotspots capture \emph{Security}~(RQ2.4).

\paragraph{Statistical Testing.}
To account for the repeated measurements of the same projects across the three eras, we fitted linear mixed-effects models with project as a random intercept and generation, programming language, and their interaction as fixed effects. The dependent variables were continuous density measures derived from the metrics. To reduce skewness and stabilize variance, these measures were log-transformed using $\log(1+x)$. We evaluated planned pairwise contrasts between generations (pre-LLM–LLM, pre-LLM–Agentic, and LLM–Agentic) using the fitted mixed-effects models.
As a non-parametric robustness check, we additionally performed paired Wilcoxon signed-rank tests on the original density values, with the rank-biserial correlation used to quantify the magnitude and direction of the paired effects and assess their practical significance. We interpreted the absolute magnitude of the rank-biserial correlation as negligible ($|r_{\mathrm{rb}}|<0.10$), small ($0.10\leq|r_{\mathrm{rb}}|<0.30$), moderate ($0.30\leq|r_{\mathrm{rb}}|<0.50$), or large ($|r_{\mathrm{rb}}|\geq0.50$). Similarly to RQ1, $p$-values were adjusted using the Benjamini–Hochberg procedure.

\subsubsection{RQ3. Software Evolution}
Lastly, we look into structural metrics and development activity to obtain an answer to the structural software evolution (RQ3).

\paragraph{RQ3.1. Structural Growth.}
Similar to the quality metrics, we also capture structural metrics with SonarQube. These metrics comprise \emph{NCLOC}, \emph{Functions}, \emph{Classes}, \emph{Files}, and \emph{Statements}, which characterize the size and structural organization of the analyzed projects. 

To account for the different durations of the eras, we compute the compound annual growth rate (CAGR)~\citep{Hatton2017} for each structural metric between consecutive snapshots. Specifically, for a metric $m$, the annualized growth between two snapshots is calculated as:

\[
\mathrm{CAGR}(m) =
\left(\frac{X_{\mathrm{end}}}{X_{\mathrm{start}}}\right)^{1/\Delta t} - 1,
\]
where $\Delta t$ is the number of years between the two snapshots. Since the initial state of each repository is assumed to contain zero structural elements, the CAGR for the initial–pre-LLM era is undefined and is therefore excluded from the analysis. We consequently obtain one project-level annualized growth rate for each structural metric for the LLM and Agentic eras. This normalization allows structural growth to be compared across these eras despite their different durations.

\paragraph{RQ3.2. Development Activity}
To characterize recent development activity, we analyze the commits and code changes made, in the default branch, during the three months preceding each era. We use a fixed three-month window to avoid longer observation windows, as they may capture changes unrelated to the development era under study (e.g., technological migrations). We report the number of commits, lines added, and lines removed for each era considering the corresponding daily rates, which is calculated as $\mathrm{DR} = \frac{v}{90}$, where $v$ is either the total number of commits, lines added, or lines removed.

\paragraph{Statistical Testing.}
For structural growth, we treat the project as the unit of analysis and compare its CAGR and daily rates between the LLM and Agentic eras using a paired two-sided Wilcoxon signed-rank test. For each comparison, we report the median values, Hodges–Lehmann estimate of the paired difference, rank-biserial effect size, and Benjamini–Hochberg adjusted $p$-value. We interpret the absolute rank-biserial correlation as negligible ($|r_{\mathrm{rb}}|<0.10$), small ($0.10\leq|r_{\mathrm{rb}}|<0.30$), moderate ($0.30\leq|r_{\mathrm{rb}}|<0.50$), or large ($|r_{\mathrm{rb}}|\geq0.50$).
\section{Results}\label{sec:results}
In this section, we present the results for our empirical analysis. The replication package with the full results is available in our online repository~\citep{EMSEGenAIReplication}.

\subsection{RQ1. Code Similarity Results} 

\paragraph{RQ1.1. Semantic Similarity.}

\begin{figure}[t]
    \centering    \includegraphics[width=1\linewidth]{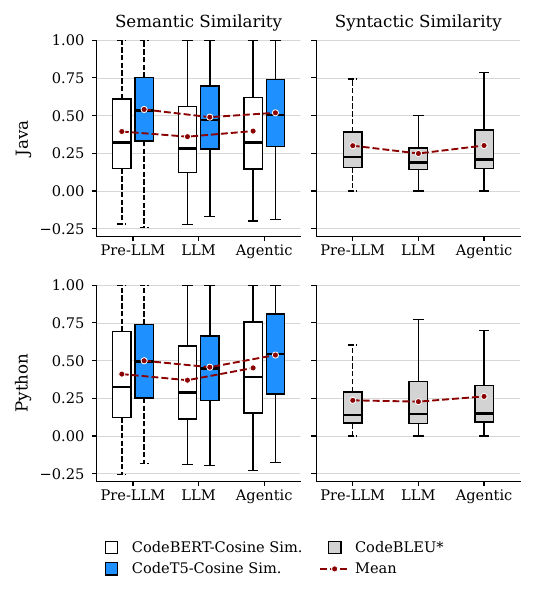}
    \caption{Cosine similarity and CodeBLEU* results for all function pairs}
    \label{fig:similarity}
\end{figure}

\figref{fig:similarity}, left side, presents the results for semantic similarity based on cosine similarity computed using the fine-tuned CodeBERT and CodeT5 models. 
For CodeBERT, the median similarity in Java decreases from 0.326 in the Pre-LLM era to 0.282 in the LLM era, and then increases to 0.324 in the Agentic era, approaching the Pre-LLM level. A similar pattern is observed for Python, where the median decreases from 0.328 to 0.290 and subsequently increases to 0.392, resulting in a higher similarity than in the Pre-LLM era. CodeT5 shows the same general pattern. In Java, the median decreases from 0.535 in the Pre-LLM era to 0.472 in the LLM era and then increases to 0.506 in the Agentic era, while in Python it decreases from 0.494 to 0.448 before increasing to 0.544. 
Thus, both semantic similarity measures exhibit the same broad directional pattern across the three development eras, although the magnitude of the changes differs across programming languages and models.

The distributions exhibit substantial variability within each development era. For CodeBERT, the middle 50\% of Java similarity scores spans 0.149--0.619 in the Pre-LLM era, 0.122--0.565 in the LLM era, and 0.145--0.630 in the Agentic era, with standard deviations of 0.303, 0.294, and 0.309, respectively. For Python, the corresponding interquartile ranges are 0.123--0.696, 0.114--0.605, and 0.153--0.760, with standard deviations of 0.333, 0.310, and 0.334. CodeT5 exhibits comparable variability. These distributions indicate substantial heterogeneity in the similarity between function pairs and considerable overlap across development eras. Descriptively, both models exhibit lower aggregate similarity in the LLM era than in the Pre-LLM era, followed by higher similarity in the Agentic era.

\begin{figure}[tb]
\centering
\includegraphics[width=1\columnwidth]{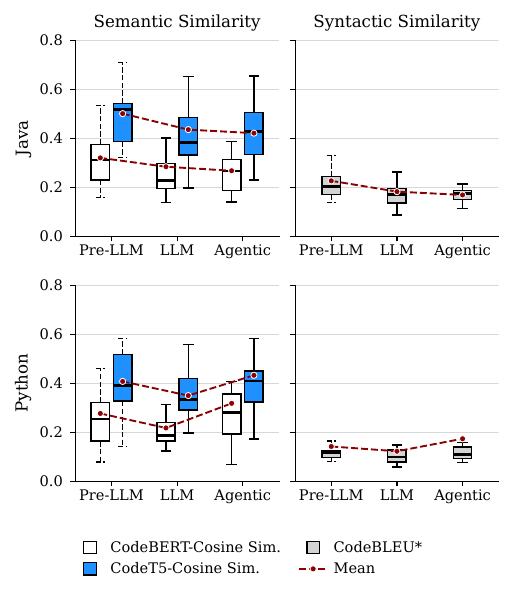}
\caption{Cosine similarity and CodeBLEU* results aggregated per project}
\label{fig:similarityProject}
\end{figure}

While  \figref{fig:similarity} includes all function pairs and is therefore influenced by projects contributing more pairs, \figref{fig:similarityProject}, on the other hand,  illustrates the project-level distributions.
Considering the statistical testing, we consider the project-level median similarities from \figref{fig:similarityProject}, giving equal weight to each project.

Statistically significant differences in semantic similarity were observed only for Java when using CodeT5. Compared with the Pre-LLM era, project-level median similarity was lower in both the LLM era ($\Delta=-0.055$, BH-adjusted $p=0.014$) and the Agentic era ($\Delta=-0.052$, BH-adjusted $p=0.014$). The corresponding paired rank-biserial correlations were $-0.696$ and $-0.725$, respectively, indicating large effects under our adopted interpretation thresholds. These results suggest that both eras exhibit greater semantic diversity than the Pre-LLM era for Java projects when measured using CodeT5. However, the difference between the LLM and Agentic eras was not statistically significant ($\Delta=-0.002$, BH-adjusted $p=0.964$), and its effect was negligible ($r_{\mathrm{rb}}=-0.014$). Thus, the project-level analysis provides no evidence of a systematic change in semantic similarity between these two eras.

For CodeBERT, neither comparison against the Pre-LLM era was statistically significant for Java, despite moderate rank-biserial correlations ($r_{\mathrm{rb}}=-0.384$ and $-0.420$ for the LLM and Agentic eras, respectively). The LLM-versus-Pre-LLM comparison for Python also exhibited a moderate effect ($r_{\mathrm{rb}}=-0.472$) but did not reach statistical significance. The remaining comparisons were likewise not statistically significant after multiple-testing correction. Overall, evidence of changes in semantic similarity is limited to Java measured using CodeT5, for which both the LLM and Agentic eras exhibit significantly lower project-level similarity than the Pre-LLM era.

\begin{leftbar}
\noindent
\textbf{RQ1.1:} Semantic similarity decreased from the Pre-LLM to the LLM era and subsequently increased in the Agentic era. These changes were statistically significant only between the Pre-LLM and the LLM era for Java measured with CodeT5. Still, both the LLM and Agentic eras showed large practical effects relative to the Pre-LLM era.
\end{leftbar}

\paragraph{RQ1.2. Syntactic Similarity.}
\figref{fig:similarity}, right side, shows the results for syntactic similarity based on the CodeBLEU* score.  
For Java, the median score decreases from 0.225 in the Pre-LLM era to 0.188 in the LLM era, and then increases to 0.212 in the Agentic era. In contrast, Python exhibits the opposite pattern, with the median increasing from 0.139 in the Pre-LLM era to 0.146 in the LLM era and 0.151 in the Agentic era. Thus, unlike the semantic similarity results, the descriptive changes in syntactic similarity differ between programming languages.

The distributions show substantial variability within each era, with standard deviations ranging from 0.181 to 0.235 for Java and from 0.218 to 0.269 for Python. The interquartile ranges also show considerable overlap across eras. Overall, the results suggest that syntactic similarity decreased from the Pre-LLM to the LLM era in Java, whereas Python shows the opposite trend.

Similar to the semantic similarity, we also report the syntactic similarity aggregated by project in \figref{fig:similarityProject}. The trends were similar, as the median (and mean) first decrease for Java projects, while they increase for Python ones. 
Moreover, the statistical analysis provides evidence of differences in syntactic similarity primarily for Java. Compared with the Pre-LLM era, both the LLM and Agentic eras exhibited significantly lower CodeBLEU* similarity, with median differences of $\Delta=-0.044$ and $\Delta=-0.040$, respectively (BH-adjusted $p=0.008$ and $p=0.001$). The corresponding rank-biserial correlations were $-0.688$ and $-0.819$, indicating \emph{large practical effects}. No significant difference was observed between the LLM and Agentic eras ($\Delta=-0.004$, $p=0.798$), with a small practical effect ($r_{\mathrm{rb}}=-0.109$). For Python, none of the comparisons reached statistical significance. However, the LLM versus Pre-LLM and Agentic versus LLM comparisons both exhibited large practical effects ($r_{\mathrm{rb}}=-0.506$ and $0.506$, respectively), despite their adjusted $p$-values of $0.063$. Thus, the statistical results support lower syntactic similarity in the LLM and Agentic eras for Java, while the corresponding Python differences were not statistically significant despite large effects in two comparisons.

\begin{leftbar}
\noindent
\textbf{RQ1.2:} Syntactic similarity decreased from the Pre-LLM to the LLM era in Java and remained lower in the Agentic era, both statistically significant. In Python, syntactic similarity increased from the Pre-LLM to the Agentic era, although this change was not statistically significant despite large practical effects.
\end{leftbar}

\begin{figure*}[!h]
\centering
\begin{subfigure}[t]{0.49\linewidth}
\begin{lstlisting}[style=pythonstyle]
def delete(self, request, *args, **kwargs):
        delete_ids = request.data['ids']
        Label.objects.filter(pk__in=delete_ids).delete()
        return Response(status=status.HTTP_204_NO_CONTENT)
        
\end{lstlisting}
\end{subfigure}
\hfill
\begin{subfigure}[t]{0.49\linewidth}
\begin{lstlisting}[style=pythonstyle]
def delete(self, request, *args, **kwargs):
        delete_ids = request.data['ids']
        self.model.objects.filter(user=request.user, 
        pk__in=delete_ids).delete()
        return Response(status=status.HTTP_204_NO_CONTENT)
\end{lstlisting}
\end{subfigure}
\vspace{-1mm}
\caption{Case 1: Pair with high semantic ($0.8$) and syntactic ($0.8$) similarity from the Pre-LLM era of project \textit{doccano}.}
\label{fig:case1}
\end{figure*}

\begin{figure*}[!h]
\centering
\begin{subfigure}[t]{0.49\linewidth}
\begin{lstlisting}[style=pythonstyle]
def _viewer_is_image_viewer(self):
        from jdaviz.configs.imviz.plugins.viewers 
        import ImvizImageView
        from jdaviz.configs.cubeviz.plugins.viewers 
        import CubevizImageView
        from jdaviz.configs.mosviz.plugins.viewers 
        import MosvizImageView, MosvizProfile2DView
        from jdaviz.configs.rampviz.plugins.viewers 
        import RampvizImageView

        def _is_image_viewer(viewer):
            return isinstance(viewer, (ImvizImageView,
            CubevizImageView, MosvizImageView,
            MosvizProfile2DView, RampvizImageView))

        viewers = self.viewer.selected_obj
        if not isinstance(viewers, list):
            viewers = [viewers]

        return np.all([_is_image_viewer(viewer) for viewer 
        in viewers])
    \end{lstlisting}
\end{subfigure}
\hfill
\begin{subfigure}[t]{0.49\linewidth}
    \begin{lstlisting}[style=pythonstyle]
def _update_available_viewers(self):
        if not hasattr(self, 'viewer'):
            return

        def is_image_viewer(viewer):
            from jdaviz.configs.imviz.plugins.viewers 
            import ImvizImageView
            from jdaviz.configs.cubeviz.plugins.viewers 
            import CubevizImageView
            from jdaviz.configs.rampviz.plugins.viewers 
            import RampvizImageView

            return isinstance(viewer, (ImvizImageView, 
            CubevizImageView, RampvizImageView))

        viewer_filter_names = [filt.__name__ for filt 
        in self.viewer.filters]
        if 'is_image_viewer' not in viewer_filter_names:
            self.viewer.add_filter(is_image_viewer)
        self.viewer_multiselect = True
        self.viewer.select_all()
    \end{lstlisting}
\end{subfigure}
\vspace{-1mm}
\caption{Case 2: Pair with high semantic (0.8) and low syntactic (0.3) similarity in the LLM era from project \textit{spacetelescope\_jdaviz}.}
\label{fig:case2}

\end{figure*}

\begin{figure*}[!h]
    \centering
    \begin{subfigure}[t]{0.51\linewidth}
        \begin{lstlisting}[style=pythonstyle]
def get_workspace_id(headers: dict, workspace_name: str) 
-> Optional[str]:

    url = f"{fabric_api_endpoint}/workspaces"

    response = requests.get(url, headers=headers)

    if response.status_code == 200:
        workspaces = response.json().get("value", [])

        for workspace in workspaces:
            if workspace.get("displayName") == workspace_name:
                return workspace.get("id")
        return None
    elif response.status_code == 404:
        print(f"[Info] No workspaces found. Response: 
        {response.text}")
        return None
    else:
        raise Exception(f"[Error] Failed to
        retrieve workspaces: {response.status_code} - 
        {response.text}")
        \end{lstlisting}
    \end{subfigure}
    \hfill
    \begin{subfigure}[t]{0.50\linewidth}
        \begin{lstlisting}[style=pythonstyle]
def get_workspace_id(workspace_name: str) -> Optional[str]:

    url = f"{fabric_api_endpoint}/workspaces"

    response = requests.get(url, headers=fabric_headers)

    if response.status_code == 404:
        print(f"[Info] No workspaces found. Response: 
        {response.text}")
        return None
    elif response.status_code != 200:
        raise Exception(f"[Error] Failed to retrieve
        workspaces: {response.status_code} -
        {response.text}")

    workspaces = response.json().get("value", [])

    for workspace in workspaces:
        if workspace.get("displayName") == workspace_name:
            return workspace.get("id")
    return None

        
        \end{lstlisting}
    \end{subfigure}
    \vspace{-1mm}
    \caption{Case 3: Pair with high semantic (0.8) and syntactic (0.8) similarity in LLM era from project \emph{Azure-Samples\_modern-data-warehouse}.}
    \label{fig:case3}
\end{figure*}

\paragraph{Analysis of representative cases.}
To complement the quantitative results, we inspect three representative function pairs illustrating different combinations of semantic and syntactic similarity across the three development eras. The complete set of results is available in our replication package~\citep{EMSEGenAIReplication}.

\paragraph{Case 1: Pre-LLM - high semantic and syntactic similarity.}
\figref{fig:case1} presents a representative function pair from the Pre-LLM era from the \textit{doccano} project. The pair obtains both a cosine similarity and a CodeBLEU* score of $0.8$. Both functions implement the same deletion workflow by obtaining the identifiers of the objects to delete from the request, filtering the corresponding objects, deleting them, and returning an HTTP 204 response.

The implementations are also structurally very similar, with the main difference being the target of the deletion operation. The function on the left operates directly on \texttt{Label}, whereas the function on the right uses \texttt{self.model} and additionally restricts the query to objects belonging to the requesting user.  Notably, these functions reside in different files, \texttt{label.py} and \texttt{comment.py}, and handle different domain entities. Nevertheless, the same deletion logic is duplicated across both files, resulting in high semantic and syntactic similarity despite their different contexts. This duplication likely resulted from implementing similar CRUD operations independently for different domain entities, rather than extracting the shared deletion logic into a reusable abstraction.

\paragraph{Case 2: LLM - high semantic but low syntactic similarity.}
\figref{fig:case2} presents a representative function pair from the \textit{spacetelescope\_jdaviz} project with a high cosine similarity of $0.8$ but a lower CodeBLEU* score of $0.3$. The pair illustrates how semantic similarity can occur at the level of a shared operation even when the overall functions implement different behaviors.

Both functions define an \texttt{is\_image\_viewer} predicate that identifies image viewers using the same \texttt{isinstance} checks for \texttt{ImvizImageView}, \texttt{CubevizImageView}, and \texttt{RampvizImageView}. However, the surrounding functionality differs substantially. The left function applies the predicate to the currently selected viewers and returns whether all selected viewers are image viewers. The right one registers the predicate as a viewer filter and subsequently enables multiselection and selects all viewers.

The first implementation additionally recognizes \texttt{MosvizImageView} and \texttt{MosvizProfile2DView}, further illustrating that the two functions are not behaviorally equivalent. Nevertheless, the shared viewer-classification logic provides a clear semantic correspondence despite differences in control flow, context, and side effects. This example demonstrates that high semantic similarity does not necessarily imply functional equivalence, as LLM-generated code may reproduce or adapt an existing semantic operation while integrating it into a different surrounding behavior.

\paragraph{Case 3: Agentic - high semantic and syntactic similarity.}
\figref{fig:case3} presents a representative function pair from the Agentic era of \textit{Azure-Samples\_modern-data-warehouse}. 
The pair obtains a cosine similarity of $0.8$ and a CodeBLEU* score of $0.8$, again indicating high semantic and syntactic similarity. 
Both functions retrieve a workspace identifier from the same API endpoint, handle unsuccessful responses, iterate over the returned workspaces, and return the identifier corresponding to the requested workspace. The implementations are highly similar, although the left function reorganizes the response-status handling and uses different names for the request headers. Notably, the two functions are located in different files (the first in ``src'', whereas the second is under ``tests''). Their close similarity and parallel functionality may indicate that one was created as a backup, test, or alternative implementation during the development process and subsequently left in the repository. Although the available evidence does not allow us to determine the exact reason for their coexistence, this example illustrates how agentic development can result in highly similar implementations being retained across different files.
\begin{figure*}[th]
    \centering
    \includegraphics[width=1\textwidth]{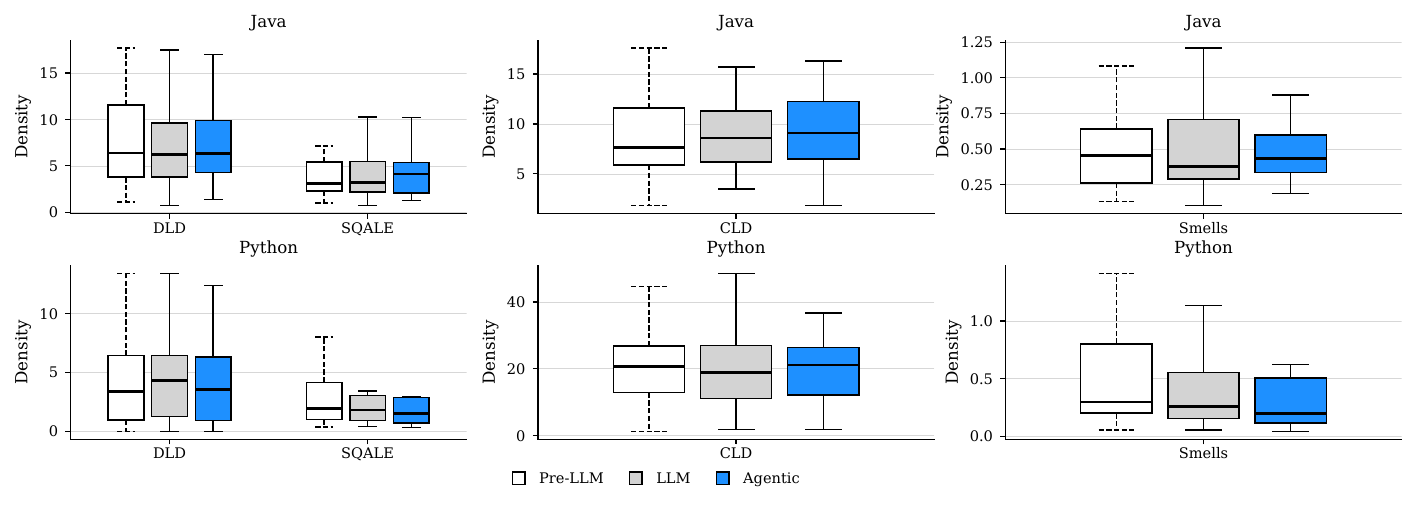}
    \caption{Maintainability results}
    \label{fig:rq21}
\end{figure*}
\begin{figure}[th]
    \centering
    \includegraphics[width=1\columnwidth]{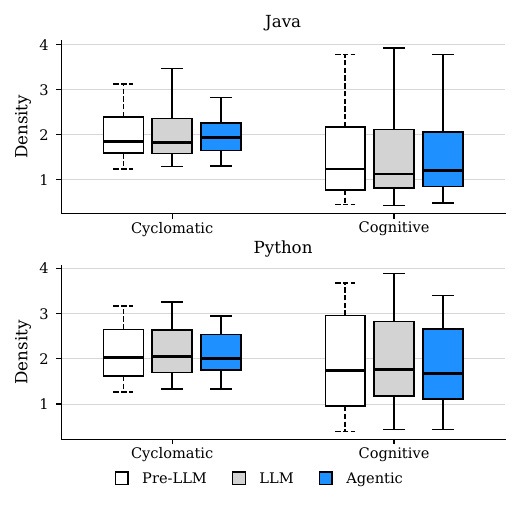}
    \caption{Complexity results}
    \label{fig:rq22}
\end{figure}
\begin{figure}[th]
    \centering
    \includegraphics[width=1\columnwidth]{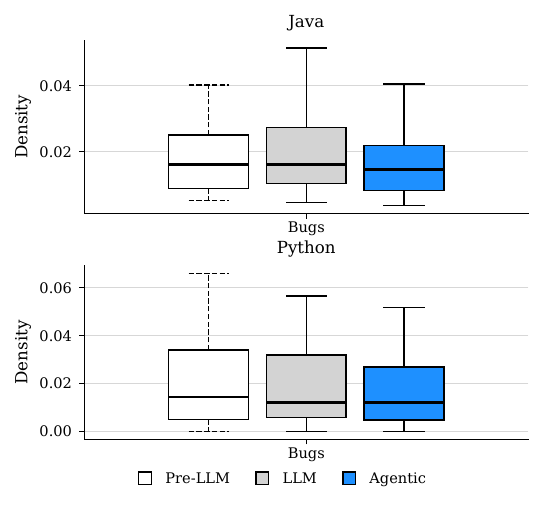}
    \caption{Reliability results}
    \label{fig:rq23}
\end{figure}
\begin{figure}[th]
    \centering
    \includegraphics[width=1\columnwidth]{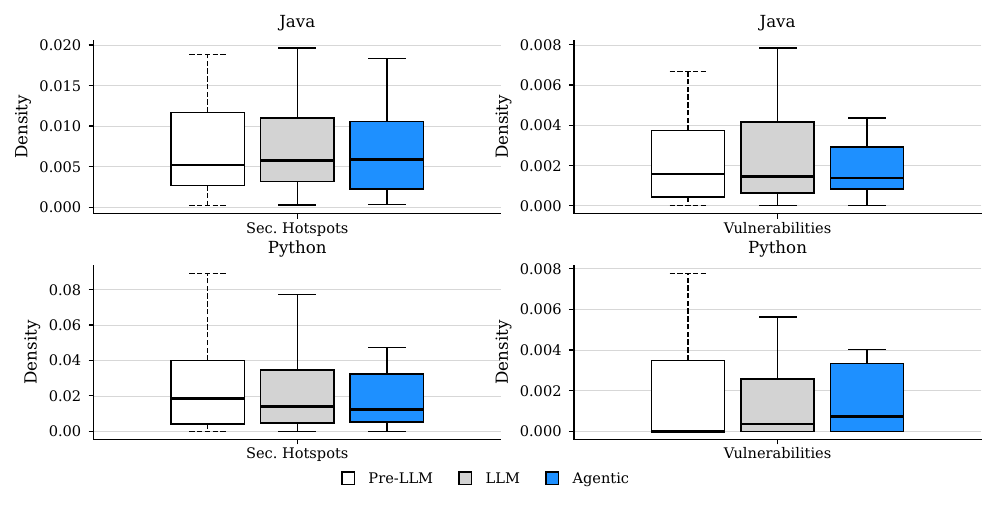}
    \caption{Security results}
    \label{fig:rq24}
\end{figure}

\subsection{RQ2. Software Quality}
\label{sec:RQ2}

In this section, we present the results for RQ2, organized according to the four quality dimensions investigated: maintainability, complexity, reliability, and security. For the quality-based metrics considered in this study, a decrease in metric density indicates an improvement in the corresponding quality dimension, whereas an increase indicates a deterioration. 

\paragraph{RQ2.1. Maintainability.}
\figref{fig:rq21} shows the distributions of the results for the Maintainability-related metrics across the three eras for Java and Python projects. Overall, Code Smell slightly decreases across the three eras in both languages, with the median decreasing from 0.453 to 0.378 in Java and from 0.298 to 0.197 in Python. SQALE Index follows a similar pattern in Python, decreasing from a median of 1.919 in the Pre-LLM era to 1.500 in the Agentic era, whereas Java shows a relatively stable distribution, with medians of 3.105, 3.212, and 4.109, respectively. DLD shows a different pattern, as Java decreases slightly from a median of 6.4 in the Pre-LLM era to 6.25 in the LLM era, before increasing to 6.35 in the Agentic era, while Python increases from 3.35 to 4.30 in the LLM era and remains at a similar level in the Agentic era (3.55). CLD remains relatively stable across eras, with Java showing a gradual increase in median from 7.65 to 9.10 and Python showing a small decrease in the LLM era followed by an increase in the Agentic era, from 20.60 to 18.95 and 21.15, respectively. Overall, the results suggest relatively small changes in maintainability across eras, with the most consistent improvement observed in Code Smell density, while the remaining metrics show language-dependent or limited changes.

\paragraph{RQ2.2. Complexity.}
\figref{fig:rq22} shows the distributions of both complexity metrics across the three eras for Java and Python projects. Overall, Java shows relatively small changes in Cyclomatic Complexity, with the median decreasing slightly from 1.841 in the Pre-LLM era to 1.822 in the LLM era before increasing to 1.930 in the Agentic era. Cognitive Complexity follows a similar pattern, decreasing from a median of 1.237 to 1.126 in the LLM era and increasing to 1.200 in the Agentic era. In Python, Cyclomatic Complexity increases from a median of 2.020 in the Pre-LLM era to 2.041 in the LLM era before decreasing slightly to 1.998 in the Agentic era. Cognitive Complexity follows the same overall pattern, increasing from 1.739 to 1.769 and then decreasing to 1.668. Overall, both complexity measures exhibit only modest changes across the three eras, with no consistent increase or decrease across languages.

\paragraph{RQ2.3. Reliability.}
\figref{fig:rq23} shows the distribution of bug density across the three eras for Java and Python projects. Considering that the number of identified bugs was small, the resulting bug density remained low, with the upper limit below 0.1. Overall, bug density remains relatively stable across the three eras in both languages. In Java, the median decreases slightly from 0.0161 in the Pre-LLM era to 0.0161 in the LLM era and 0.0145 in the Agentic era. Python shows a similar pattern, with the median decreasing from 0.0143 in the Pre-LLM era to 0.0119 in the LLM era before increasing slightly to 0.0121 in the Agentic era. The substantially higher means and maximum values observed for Java are driven by a small number of projects with exceptionally high bug densities, resulting in highly skewed distributions. Overall, the median bug density shows only limited variation across the three eras, with no consistent directional change.

\paragraph{RQ2.4. Security.}
\figref{fig:rq24} shows the distributions of the Security metrics across the three eras for Java and Python projects. Overall, vulnerability density remains low and relatively stable across the three eras. In Java, the median vulnerability density decreases from 0.00157 in the Pre-LLM era to 0.00145 in the LLM era and 0.00138 in the Agentic era. Python shows a similar low-density pattern, with the median increasing from 0 in the Pre-LLM era to 0.00036 in the LLM era and 0.00073 in the Agentic era. Security Hotspot density, however, shows greater variation between languages, with Java remaining relatively stable across eras, increasing slightly from a median of 0.00522 in the Pre-LLM era to 0.00588 in the Agentic era. In Python, the median decreases from 0.0183 in the Pre-LLM era to 0.0121 in the Agentic era. Overall, the results indicate limited changes in security-related metric densities across the three eras, although the distributions differ substantially between Java and Python.

\paragraph{Statistical testing results.}
The mixed-effects analysis identified statistically significant differences for two metrics after FDR correction. Compared with the Pre-LLM generation, the Agentic generation exhibited a $7.34\%$ lower code-smell density (95\% CI: $[-10.54\%, -4.03\%]$, $p<0.001$, $q<0.001$) and a $10.98\%$ lower SQALE index density (95\% CI: $[-18.16\%, -3.16\%]$, $p=0.0068$, $q=0.044$), evidencing improvements in these maintainability-related measures. No statistically significant generation $\times$ language interactions were detected, indicating that the observed generation effects did not differ significantly between Java and Python. The paired Wilcoxon signed-rank robustness analysis identified a statistically significant change only for code-smell density between the Pre-LLM and Agentic generations ($W=322$, $p=0.0019$, $q=0.017$), with a moderate rank-biserial effect ($r_{\mathrm{rb}}=-0.495$). Although code-smell density also showed \emph{moderate practical effects} for the Pre-LLM--LLM ($r_{\mathrm{rb}}=-0.418$) and LLM--Agentic ($r_{\mathrm{rb}}=-0.357$) comparisons, neither remained statistically significant.

\begin{leftbar}
\noindent
\textbf{RQ2:} Software quality showed limited changes across the three eras, with the most notable statistically significant differences observed in maintainability. In particular, the Agentic era exhibited lower code-smell and SQALE index densities than the Pre-LLM era, while complexity, reliability, and security metrics showed relatively small or language-dependent changes without statistically significant effects.
\end{leftbar}

\subsection{RQ3. Software Evolution}

\paragraph{RQ3.1. Structural Growth.}

\begin{figure}[th]
    \centering
    \includegraphics[width=1\columnwidth]{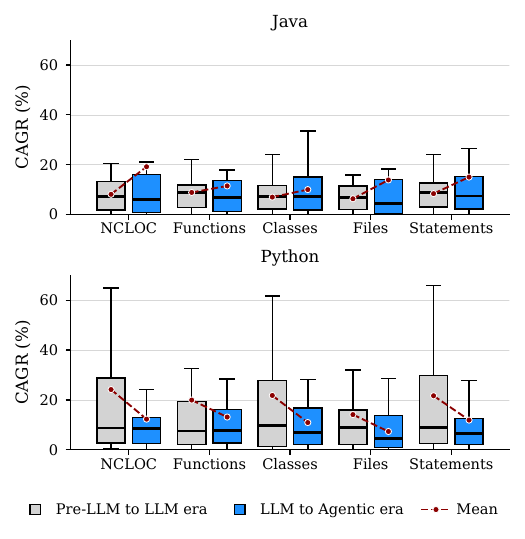}
    \caption{Structural compound annual growth rates (CAGR) results per programming language.}
    \label{fig:cagr}
\end{figure}

\figref{fig:cagr} shows the CAGR of the five structural metrics across the pre-LLM to LLM era and the LLM to agentic era. For Java projects, structural growth remained positive for most projects in both eras, although the median growth rate generally decreased in the LLM to agentic era. Median NCLOC growth decreased from 7.30\% to 6.10\%, while function growth decreased from 8.76\% to 6.80\%. File growth showed the largest reduction, from 6.86\% to 4.42\%. Statement growth also decreased, from 8.74\% to 7.45\%. In contrast, class growth remained almost unchanged, increasing slightly from 7.19\% to 7.32\%. Thus, for Java, the Agentic era is characterized primarily by slower growth (notice the mean increase) in lines of code, functions, files, and statements, while the growth rate of classes remains similar.

Python projects exhibit a somewhat different pattern (notice the mean decrease). Median NCLOC growth decreases only slightly, from 8.69\% in the pre-LLM to LLM era to 8.48\% in the LLM to agentic era. Function growth is the only metric whose median increases, from 7.44\% to 7.73\%. In contrast, class growth decreases from 9.63\% to 6.83\%, file growth decreases from 8.96\% to 4.41\%, and statement growth decreases from 8.82\% to 6.37\%. The reduction is therefore particularly pronounced for classes, files, and statements.

The distributions also show substantial project-level variability. For Java, for example, Classes CAGR goes up to $\approx$35\% in the LLM-to-Agentic era. Python projects, however, exhibit positive minimum growth rates in the first era. The wide ranges indicate that the median differences do not represent a uniform change across all projects.

Despite these changes, none of the differences was statistically significant after correction for multiple comparisons, with adjusted $p$-values ranging from $0.4277$ to $0.9238$. The rank-biserial effect sizes were negative for all metrics, ranging from $-0.0165$ for Functions (negligible) to $-0.2376$ for Files (small), indicating generally small practical differences. Thus, although structural growth was consistently lower in the Agentic era at the median level, the observed differences were not sufficiently consistent across projects to establish a statistically significant change in structural growth.

\begin{leftbar}
\noindent
\textbf{RQ3.1:} The LLM-to-agentic era shows slower growth, especially for Python, across all five structural metrics, with the largest reduction observed for files. These differences, however, are not statistically significant.
\end{leftbar}

\paragraph{RQ3.2. Development Activity.}
\begin{figure}[th]
    \centering
    \includegraphics[width=.8\linewidth]{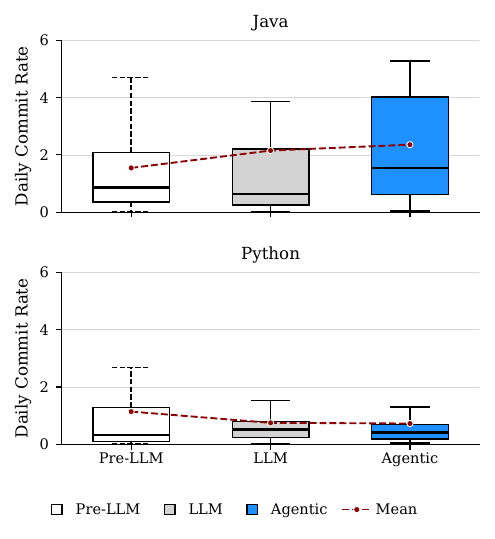}
    \caption{Daily commit rate during the three-month period preceding each snapshot era.}
    \label{fig:dcr}
\end{figure}
\begin{figure}[th]
    \centering
    \includegraphics[width=.9\linewidth]{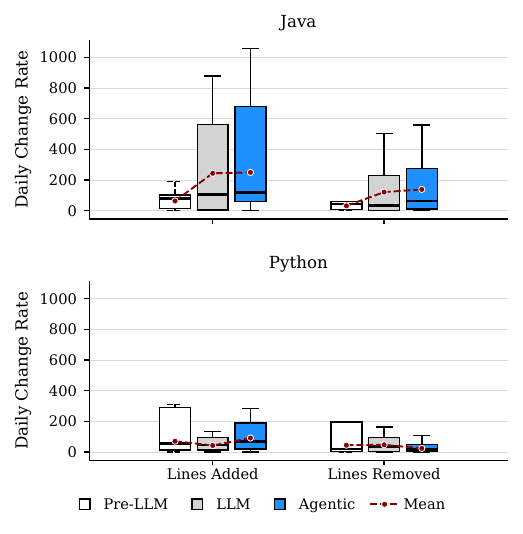}
    \caption{Development activity in terms of daily lines added and removed across development eras.}
    \label{fig:devActivity}
\end{figure}

\figref{fig:dcr} shows the distribution of the daily commit rate across the three eras, considering the three-month period preceding each snapshot. As illustrated, Java projects exhibit a decrease in median from $0.8556$ in the Pre-LLM era to $0.6333$ in the LLM era, followed by a substantial increase to $1.5444$ in the Agentic era. A similar pattern is observed considering the upper quartile of the distributions.
Python projects, on the other hand, show the median increasing from $0.3333$ in the Pre-LLM era to $0.5111$ in the LLM era before slightly decreasing to $0.4222$ in the Agentic era. The distributions also remain highly variable across projects, with particularly large ranges in the Java and Python Pre-LLM eras.

Considering the statistical testing results, Java projects showed an increase with a Hodges--Lehmann estimate of $0.2000$ and a moderate rank-biserial effect size of $0.4708$. This difference, however, was not statistically significant  ($p=0.0787$). Python projects had a Hodges--Lehmann estimate of $0.0333$ and a negligible rank-biserial effect size of $-0.0123$ ($p=0.9571$). Thus, although Java projects exhibited a moderate increase in development activity in the Agentic era, the evidence was insufficient to establish a statistically significant change, while Python projects showed essentially no difference.

To complement these results, we also analyze the content of these commits (i.e., lines added and removed). \figref{fig:devActivity} illustrates these values, with different patterns for Java and Python. For Java projects, median daily lines added increase from 78.7 in the Pre-LLM era to 105 in the LLM era and 118.9 in the Agentic era (the mean follows a similar trend). The median for the daily removed lines decreases slightly from 43.4 to 35 in the LLM era, before increasing to 62.8 in the Agentic era. For Python, lines added initially decrease from 55.2 to 45.6 in the LLM era, then increase to 68 in the Agentic era. Daily lines removed increase from 19.8 to 33.9 in the LLM era and subsequently decrease to 14.6 in the Agentic era. These results suggest that agentic development is associated with higher code addition activity in both languages, whereas code removal follows different patterns across Java and Python. The large difference in the interquartile range also indicates a substantial difference in development activity across projects, particularly for Java, with a max of one thousand lines added per day during the agentic era (outliers excluded).

Considering the statistical testing, Java projects showed a moderate effect for the increase in lines added daily from the Pre-LLM to the Agentic era (rank-biserial $=0.3143$), although this difference was not statistically significant ($p=0.6652$). Python projects showed a moderate effect for the decrease in lines added from the Pre-LLM to the LLM era (rank-biserial $=-0.4308$), which was also not statistically significant ($p=0.6652$). The remaining comparisons showed only small or negligible effects, including the transition from the LLM to Agentic era for both languages.

\begin{leftbar}
\noindent
\textbf{RQ3.2:} Development activity increased moderately for Java projects in the Agentic era, particularly in commit and lines-added rates, while Python projects showed little change in overall activity. Despite these trends, the differences were not statistically significant.
\end{leftbar} 
\section{Discussion}
\label{sec:discussion}
In this section, we present an analysis of the findings, highlighting the key takeaways from the results. 
 
\subsection{Code Similarity, Diversity, and Predictability}

\paragraph{Similarity trends across development eras.}
RQ1 does not reveal a uniform trend toward greater code similarity following the adoption of LLMs and agentic AI. Instead, the observed changes vary by programming language and similarity measure. For Java, both CodeT5-based semantic similarity and CodeBLEU* syntactic similarity are significantly lower in the LLM and Agentic eras than in the Pre-LLM era, with large effect sizes. In contrast, Python exhibits less consistent patterns as semantic similarity increases in some Agentic-era comparisons, while syntactic similarity shows no statistically significant differences across eras. These findings challenge the assumption that AI-assisted development necessarily leads to greater code homogeneity and indicate that its relationship with code similarity depends on the language and representation considered.

One possible explanation is that the two languages serve different application domains and impose different programming conventions. Java is frequently used in enterprise systems~\citep{Farley2006}, where explicit typing and more structured implementations may shape how AI-generated code is integrated into existing codebases. Python, in contrast, is widely used across diverse domains, including data science and scripting, and its flexible syntax may allow more varied implementations of similar functionality~\citep{Peng2021}. Differences in the adoption and use of AI coding tools across these ecosystems may also contribute to the observed patterns. These assumptions, however, were not investigated as part of this study and remain as future research opportunities.

\begin{leftbar}
\noindent \textbf{Key takeaway 1:} AI-assisted development does not necessarily produce uniform similarity trends across languages and representations.
\end{leftbar}

\paragraph{Predictability and software ecosystem evolution.}
LLMs learn statistical regularities from existing code~\citep{Hindle2012Naturalness,Allamanis2018BigCode}, which can help them generate code that follows common programming patterns. However, prior work has also raised concerns about the reproduction of existing code, including training-data memorization and the generation of code similar to open-source implementations~\citep{Wan2024DoesYN,Guo2025LLMsAB,Xu2024LicoevalEL}. These findings motivate the expectation that AI-assisted development could increase code similarity.

Our results, however, do not support this expectation uniformly, as semantic and syntactic similarity decreased significantly for Java, while Python exhibited different trends depending on the similarity measure. This contrast suggests that the relationship between AI-assisted development and code similarity is more nuanced than a general tendency toward homogenization. These trends, however, may also be a reflection of the natural evolution of software systems over time, as repeated refactoring and maintenance can alter code structure and reduce similarity between implementations. Nevertheless, when considered alongside the codebase growth observed in RQ3.1, the decline in similarity suggests that software growth does not necessarily lead to greater code homogeneity.

\begin{leftbar}
\noindent \textbf{Key takeaway 2:} Our results challenge the assumption that AI-assisted development necessarily homogenizes code.
\end{leftbar}

\paragraph{Contextual interpretation of code similarity.}
The representative cases discussed in the RQ1 results illustrate how similarity reflects different development contexts. 
In this sense, the context in which AI-generated code is produced may also influence these patterns. Prompts or repository-level instructions, such as \texttt{AGENTS.md} files, can encourage consistency with existing conventions, reuse of established abstractions, and adherence to project-specific architecture. Such guidance may make generated code more predictable and compatible with a project, but it could also reinforce existing patterns (i.e., making software less diverse). Thus, instructions that encourage different designs or discourage duplicated code may also affect similarity. Although we did not study these instructions directly, they are worth considering when examining how AI adoption affects code similarity.

\begin{leftbar}
\noindent \textbf{Key takeaway 3:} Prompts and repository-level instructions may influence how LLMs and agents reuse existing code and follow project conventions, thereby shaping code similarity and diversity.
\end{leftbar}

\subsection{Software Quality and its Impact}

\paragraph{Overall quality trends.}
RQ2 results provide little evidence of systematic software quality deterioration following the adoption of LLMs and agentic AI. Moreover, the mixed-effects models identify no statistically significant generation-era-by-language interactions, providing insufficient evidence that the overall era-related patterns differ systematically between Java and Python. These findings challenge the assumption that the increasing use of AI-assisted development necessarily degrades the quality of long-lived open-source projects.

 \begin{leftbar}
\noindent \textbf{Key takeaway 4:} LLM- and agent-assisted development is not associated with consistent software quality deterioration or systematic differences between Java and Python.
\end{leftbar}

\paragraph{Maintainability-related indicators.}
The mixed-effects analysis identifies two significant differences after multiple-testing correction: code-smell density decreases by 7.34\% and SQALE density by 10.98\% in the Agentic era relative to the Pre-LLM era. The paired analysis independently confirms a reduction in code-smell density, with a moderate effect size. These findings indicate improvements in the measured maintainability-related indicators, although they cannot be attributed directly to AI adoption. Other factors, such as refactoring, changes in development practices, and improvements in static analysis tools over time~\citep{Vassallo2020}, may also contribute to the observed trends.

\begin{leftbar}
\noindent \textbf{Key takeaway 5:} LLM- and agent-assisted development coincides with improvements in measured maintainability-related indicators.
\end{leftbar}

\paragraph{Security and other quality indicators.}
The remaining metrics show no consistent changes across development eras. Complexity remains relatively modest, while bug-related density is low and broadly stable. Security indicators vary by language: Python security-hotspot density decreases between the Pre-LLM and Agentic eras, whereas Java shows a small increase. Vulnerability density decreases slightly in Java but rises from zero in the Pre-LLM snapshot to a small positive value in the Python Agentic snapshot. These results require careful interpretation because security hotspots indicate code requiring review rather than confirmed vulnerabilities. Moreover, changes in detected issues may reflect differences in the analyzed code or in the coverage of static-analysis rules.

\begin{leftbar}
\noindent \textbf{Key takeaway 6:} LLM- and agent-assisted development shows no clear impact on security or complexity quality indicators.
\end{leftbar}

\paragraph{Implications for AI-assisted development.}
Taken together, our findings suggest that the increasing adoption of LLMs and agentic AI is not associated with broad deterioration in the software quality indicators examined, although neither do they establish an overall improvement. While code-smell and SQALE densities decrease in the Agentic era, complexity and bug-related densities remain broadly stable, and security indicators show mixed trends across languages. Similarly, the observed differences between semantic and syntactic similarity, particularly across Java and Python, demonstrate that conclusions about code evolution depend on the representations and measures considered. These findings highlight the need to evaluate AI-assisted software evolution using complementary similarity measures and multiple quality dimensions rather than relying on individual indicators. 

Combining similarity measures with function roles, repository structure, and development history could help distinguish code reuse, duplicated functionality, architectural variation, and independent implementations. Furthermore, future studies should investigate whether changes in static-analysis indicators correspond to improvements in functional correctness, test effectiveness, security, and long-term maintainability, and whether these changes persist as software is integrated, maintained, and evolves.

\begin{leftbar}
\noindent \textbf{Key takeaway 7:} LLM- and agent-assisted development shows encouraging maintainability trends but no consistent improvement across all quality dimensions. Hence, evaluating software evolution requires complementary similarity measures, as individual indicators do not necessarily imply improvements in quality.
\end{leftbar}

\subsection{Structural Growth and Development Activity.}

\paragraph{Structural growth across development eras.}
The structural-growth analysis reveals no statistically significant changes in CAGR between the Pre-LLM, LLM, and Agentic eras after multiple-testing correction. Across Java and Python, median growth rates for most measures are lower in the LLM-to-Agentic interval than in the Pre-LLM-to-LLM interval, although Java class growth and Python function growth increase slightly. The effect sizes are generally negligible to small, providing insufficient evidence that structural growth has systematically changed across the sampled projects.

\begin{leftbar}
\noindent \textbf{Key takeaway 8:} LLM- and agent-assisted development is not associated with changes in structural growth rates.
\end{leftbar}

\paragraph{Development activity across development eras.}
Development activity exhibits different trends across languages. Although the increase from the Pre-LLM to the Agentic era has a moderate effect size in Java projects, it does not remain statistically significant. Python exhibits a different pattern, with a median decrease across the three eras, respectively, and no significant differences. Code changes in terms of added/removed also vary by language: additions and removals both increase in the Agentic era for Java, whereas Python shows increased additions but decreased removals relative to the LLM era.

These differences in development activity can be considered alongside the similarity trends discussed earlier, which also vary between Java and Python. Java's lower similarity in the LLM and Agentic eras (RQ1), together with increased additions and removals in the Agentic era (RQ3.2), may reflect more extensive changes to existing implementations. In contrast, Python's less consistent similarity trends (RQ1), combined with increased additions but decreased removals relative to the LLM era (RQ3), may reflect different patterns of code evolution. Nonetheless, these differences may be related to the languages' typical application domains, programming conventions, and AI adoption practices. Investigating if these assumptions hold remains as future work.

\begin{leftbar}
\noindent \textbf{Key takeaway 9:} LLM- and agent-assisted development shows different activity patterns across Java and Python, with a descriptive increase in Java commit rates in the Agentic era.
\end{leftbar}

\paragraph{Interpreting structural growth and development activity.}
Structural growth and development activity capture different aspects of software evolution. More commits or more lines changed do not necessarily imply faster structural growth, since development activity may include refactoring, maintenance, test changes, and dependency updates that do not increase the measured structure~\citep{Fowler2002}. Software modernization, such as updating libraries or migrating to new versions, can also require substantial code changes~\citep{Assuncao2025}. Conversely, lower growth rates do not necessarily indicate reduced effort or productivity. The increase in Java activity during the Agentic era may reflect intensified development or maintenance, although our results do not establish its cause.

Differences across languages may also reflect project-specific practices and architectural characteristics. Python's flexible typing and dynamic object behavior may allow some changes to be implemented with fewer code modifications, whereas Java's static typing and explicit class structures may require more extensive changes when modifying interfaces or evolving system architecture.
Taken together, our results provide no conclusive evidence that the transition to LLM-assisted or agentic development has systematically accelerated or slowed the structural evolution of long-lived projects. Although the descriptive trends identify changes worth investigating, the lack of statistically significant differences limits stronger conclusions. Future analyses could use finer-grained development histories to distinguish newly introduced functionality from refactoring, maintenance, generated-code integration, and other types of change.

\begin{leftbar}
\noindent \textbf{Key takeaway 10:} Changes in activity during LLM- and agent-assisted development do not necessarily translate into faster or slower structural growth.
\end{leftbar}

\paragraph{Understanding repository context and development practices.}
Repository context and developer practices need further investigation to understand how AI-assisted development affects code similarity and quality. Agentic systems can use project-specific instructions, existing abstractions, tests, and architectural constraints to guide code generation, potentially promoting consistency while also reinforcing existing patterns.

Recent studies have begun to examine these interactions in practice by analyzing open-source coding conversations and repository histories around AI adoption, finding no deterioration in bug-fix, test-related, or CI-related code indicators~\citep{Fang2026Conversation}. Similarly, \citet{Cynthia2026BeyondBugFixes} compare base and merged revisions of agent-generated Python bug-fix pull requests using SonarQube, showing that successful merging does not reliably indicate acceptable post-merge code quality. Although these studies provide evidence on code quality and repository evolution, they do not fully explain how repository context and developer practices shape the similarity and quality of AI-generated contributions over time. Furthermore, recent developments in industry~\citep{Anthropic2026} have raised concerns about the limitations of existing version control systems, such as Git, in supporting agent-driven software development. These concerns may signal a shift toward alternative version control technologies and paradigms better suited to this development model. Future research should therefore investigate how developers review, modify, and integrate AI-generated contributions, considering not only the resulting artifacts and the version control strategies used to manage them, but also the processes through which these contributions are incorporated into existing software systems.

\begin{leftbar}
\noindent \textbf{Key takeaway 11:} Future work should investigate how developers review, integrate, and maintain AI-generated code, and how version control systems and development workflows must evolve to support agent-driven development.
\end{leftbar}

\section{Limitations and Threats to Validity}
\label{sec:threats}
In this section, we present the limitations and threats to the validity, organized around the internal, external, construct, and conclusion validity categories reviewed by \citet{feldt2010validity}.

\subsection{Internal Validity}

\emph{Temporal comparisons and alternative explanations.}
Comparing the same projects across three eras reduces the influence of stable characteristics such as domain, architecture, and conventions. However, observed changes in similarity, quality, structural growth, and development activity cannot be attributed exclusively to LLMs or agentic AI. Factors such as project aging, framework migrations, changing requirements, contributor turnover, and evolving practices may also contribute. All projects show traces of agentic adoption and there is no contemporaneous non-adopting group, so AI's contribution cannot be isolated. In addition, the calendar-based snapshots add a limitation as the 2021--2024 interval includes pre-ChatGPT time, and the 2026 snapshot does not imply equivalent duration or intensity of agent use across projects. Likewise, the 90-day windows in RQ3.2 and the annualization of structural changes in RQ3.1 standardize duration but not AI exposure.

\emph{Differences in the function populations.}
The RQ1 populations differ in size and history. The Pre-LLM baseline contains accumulated functions from the 2021 snapshot, whereas the LLM and Agentic populations contain functions newly introduced relative to the preceding observation and retained at the endpoint. The diff filter retains approximately 100K (LLM) and 90K (Agentic) functions versus approximately 1M in the baseline, so these populations may differ in task composition, history, and survival likelihood, and a larger candidate pool offers more chances to find highly similar functions even if the process is unchanged. We limit computational cost by retrieving the top $k=20$ nearest neighbors per function, selected through a validation study based on exhaustive pairwise comparison of a repository subset, but fixing $k$ does not equalize pool sizes. The larger baseline thus remains a potential source of bias in baseline-to-later comparisons.

\emph{Function matching and snapshot survival.}
The diff filter identifies new functions by signature, which scales well but may misclassify changes. Java matching uses the class name, method name, and parameter names and types, whereas Python matching uses the file name, function name, and number of parameters due to parser limitations, creating different opportunities for missed matches and identity collisions. Signature-preserving modifications, including potentially extensive AI-assisted rewrites, may be excluded, while renamed or moved functions may be classified as new. Since the same rules apply across eras within each language, these limitations are unlikely to invalidate the broader comparisons.

Non-behavior-relevant functions are another threat, which we mitigate by excluding short functions and boilerplate methods (getters, setters, entry points) based on length, token count, and name. Finally, snapshots cannot capture functions introduced and later removed, or reverted and revised changes, so our conclusions concern functions retained at the selected snapshots and identified as new under our extraction rules, not the complete development history.

\subsection{External Validity}

\emph{Project eligibility and sample representativeness.}
Project selection is a relevant threat. Projects had to be created before 2021, have at least 1,000 commits, remain active since the start of the LLM era, involve multiple contributors, and show engineering practices through a license, a CI/CD descriptor, and a README. We also exclude shared-commit forks and require at least ten files and snapshots in all three target years. These criteria focus on mature, collaborative projects while excluding younger, abandoned, non-adopting, and private projects, and those without recognized agent traces. We used stratified random sampling to select 25 Java and 25 Python projects from 912 eligible ones; Kolmogorov--Smirnov tests on commit counts, distinct authors, and file counts show no significant differences between each sample and its stratum. The equal language allocation, however, does not reflect the eligible population's proportions, so findings should generalize primarily to projects meeting our criteria rather than the entire open-source ecosystem.

\emph{Software Heritage coverage.}
The sampling frame includes 13,123 of the 27,192 repositories in the source agent-adoption dataset that were present in the Software Heritage export dated 2026-06-04. Because archive coverage is incomplete and ingestion can be delayed, missing repositories may introduce selection bias. We sampled 1,000 absent repositories and retrieved their creation dates from GitHub: of the 983 still accessible, 87.3\% were created after 2021-01-01 and would have been excluded by our criteria. This supports the expectation that many missing repositories would not qualify, but does not show that the remainder, particularly older projects meeting the activity and engineering criteria, are comparable to those included, so archive-related bias remains possible.

\emph{Transferability across languages and development settings.}
Finally, the study covers only Java and Python projects with detectable agentic-development traces. Language-specific extraction and matching, differences in static-analysis coverage, and the chosen eras may limit transferability to other languages, project sizes, workflows, and governance settings, and findings may change as agents, models, and practices evolve. Replications with additional languages, projects without detectable adoption traces, and independently verified AI-exposure histories would clarify how far the patterns generalize.

\subsection{Construct Validity}

\emph{AI adoption and development eras.}
The eras rely on the agent-adoption dataset of \citet{Robbes2026AgenticMA}, which identifies usage through heuristics over configuration files, commit authorship and metadata, and collaborative workflows. These indicators do not quantify how intensively agents were used or establish that a given function was agent-generated or modified. 
Classifying 2021 as Pre-LLM does not rule out earlier automated assistance, the 2024 snapshot reflects potentially LLM-assisted rather than verified LLM-exposed development, and the 2026 snapshot contains both agent and non-agent authored code. Although causal effects are not isolated, the comparisons provide empirical evidence of how software characteristics change alongside AI-assisted development.

\emph{Operationalization of code similarity.}
RQ1 measures semantic similarity via cosine similarity from CodeBERT and CodeT5 models fine-tuned for Type-IV clone detection~\citep{Marchezan2026Kamino}, and syntactic similarity via CodeBLEU*. These metrics do not establish functional equivalence or the diversity of solutions to a fixed problem, but capture complementary dimensions; fine-tuning for clone detection makes the embeddings more relevant to semantic relatedness, and two distinct models show whether patterns hold across representations. CodeBLEU* averages lexical, syntactic, and keyword-matching components with the data-flow component disabled, so it does not capture all aspects of program semantics. Candidate pairs for all three measures come from pretrained CodeBERT embeddings and an approximate nearest-neighbor index, with $k=20$ chosen via Recall@$k$ against an exhaustive reference set of clone pairs in a repository subset. This supports the retrieval strategy but does not guarantee coverage of all possible similarities, so our findings characterize similarity among retrieved candidates, though the number of pairs remained large.

\emph{Within-project diversity and code scope.}
RQ1 measures similarity among selected functions within each project, not architectural originality, cross-project convergence, or diversity of implementations of the same requirement. Distributions may shift because the task composition changes even if developers' approaches remain unchanged. Notebook-based and script-based Python development may also differ~\citep{grotov2022large}; we mitigated this by creating functions only from \emph{.py} files.

\emph{Software quality metrics and normalization.}
RQ2 uses SonarQube Community Edition metrics for maintainability, complexity, reliability, and security. These are consistent, tool-based indicators, but do not establish actual failures or maintenance effort: code smells are static-analysis findings rather than confirmed problems, the SQALE index estimates remediation effort rather than actual effort, bugs and vulnerabilities are analyzer findings, and security hotspots merely flag code for review. The low counts of bugs and vulnerabilities further constrain reliability and security conclusions. Size-normalized metrics, within-project comparisons, and complementary analyses (e.g., mixed-effects models) nonetheless provide a systematic basis for identifying changes in static-analysis indicators while accounting for project characteristics and metric distributions.

We normalize metrics by the number of functions (density) to compare projects of different sizes, but changes in function counts may also affect densities and should be considered when interpreting trends. DLD and CLD retain their native SonarQube definitions and use different denominators, so we analyze each measure independently across eras rather than comparing absolute values across metrics. Although analyzer rules and coverage may vary between Java and Python, consistent analyzer versions, rule profiles, exclusions, and function-count criteria across snapshots support reproducible within-project comparisons.

\emph{Structural growth and development activity.}
RQ3 measures structural growth via CAGR of non-comment lines of code, functions, classes, files, and statements, and activity via daily rates of commits, added lines, and removed lines in the 90 days preceding each snapshot. These do not directly measure productivity, effort, or change quality: a higher commit rate may reflect smaller commits, increased code volume may stem from generated code, new functionality, or style changes, and added and removed lines do not distinguish refactoring from churn. Fixed windows and CAGR make measures comparable across durations and intervals but do not control for the nature of the work, and CAGR captures only net change between endpoints. Results should thus be read as evidence of structural evolution and recorded activity, not of productivity or efficiency.

\emph{Illustrative function pairs.}
The RQ1 cases illustrate possible forms of semantic and syntactic similarity, such as duplicated operations across files and implementations in different contexts. They are selected, not random or exhaustive, and cannot establish prevalence, developer intent, or LLM or agent provenance; we use them only to contextualize the quantitative results.

\subsection{Conclusion Validity}

\emph{Statistical power and units of analysis.}
The many function pairs in RQ1 are not independent observations, as pairs share functions and projects. We therefore aggregate pair-level scores using the median per project, era, language, and similarity measure, and run paired Wilcoxon signed-rank tests with projects as the units of analysis. RQ2 uses linear mixed-effects models with project-specific random intercepts, and RQ3 likewise relies on project-level comparisons. Nevertheless, the effective sample is 50 projects (25 per language), limiting sensitivity to small effects and language-specific differences, particularly for reliability and security metrics, where few bugs and vulnerabilities were observed. The Cochran-based sample-size calculation supports precision for estimating a proportion, but does not establish power for the paired tests or mixed-effects models. Non-significant findings should thus be read as potential trends, not as evidence of equivalence or absence of an effect.

\emph{Aggregation, multiple comparisons, and model assumptions.}
Finally, we report effect sizes alongside significance and apply the Benjamini--Hochberg procedure within the specified test families, which reduces, but does not eliminate, sensitivity to the chosen procedures. Project-level medians give each project equal weight and limit the influence of extreme scores, but can overlook changes concentrated in a few functions or in the highly similar tail; descriptive pair-level distributions also reflect unequal numbers of pairs per project. In RQ2, the $\log(1+x)$ transformation reduces skewness and stabilizes variance, yet the mixed-effects models retain distributional and specification assumptions, complemented by Wilcoxon tests that still require careful interpretation. The robustness analysis corroborates the code-smell finding. Finally, differences in significance between Java and Python should not alone be interpreted as a language effect.

\section{Related work}\label{sec:related}
In this section, we present work related to our study, which is summarized in \tabref{tab:closest-related-work}. 

\begin{table*}[t]
\caption{Closest empirical designs. ``Measures'' summarizes the main outcomes or variables analyzed in each study. ``Similarity'' refers to function/output similarity.}
\label{tab:closest-related-work}
\centering
\small
\setlength{\tabcolsep}{3pt}
\renewcommand{\arraystretch}{1.15}
\begin{tabularx}{\linewidth}{
@{}
>{\raggedright\arraybackslash}p{0.18\linewidth}
>{\raggedright\arraybackslash}X
>{\raggedright\arraybackslash}X
>{\raggedright\arraybackslash}p{0.49\linewidth}
@{}}
\toprule
\textbf{Study} & \textbf{Material and time unit} & \textbf{AI evidence needed} & \textbf{Measures} \\
\midrule
\citet{Xu2026CodeTransformed}
    & Research-linked OSS, 2020--2025 corpus
    & No function labels for temporal corpus
    & Naming patterns and identifier length; cyclomatic and Halstead complexity; SLOC, comment ratio, and maintainability index; cosine/Jaccard code similarity; reasoning-label match/error
    \\

\citet{He2026SpeedCostQuality}
    & OSS histories around Cursor adoption
    & Observable adoption proxy
    & Monthly commits and lines added; static-analysis warnings, duplicated-line density, and cognitive complexity
    \\

\citet{Agarwal2026AIIDEsAutonomousAgents}
    & Monthly OSS outcomes around agents
    & Agent PR traces; prior IDE signals
    & Monthly commits and lines added; static-analysis warnings, duplicated-line density, comment-line density, and cognitive complexity
    \\

\citet{Fang2026Conversation}
    & Shared chats and monthly OSS histories
    & Shared coding conversations
    & AI-related commit share and churn; chat purposes; commits; bug/fix and bug-issue signals; test-touching commits and CI outcomes; issue/PR activity and resolution times; contributor/review participation and concentration; survey perceptions
    \\

\citet{Shi2025}
    & Machine- and human-authored code
    & Known code provenance
    & Lexical similarity (token frequency, syntax-element distributions, Zipf/Heaps); conciseness (tokens and lines); naturalness (token log-likelihood/rank); detector performance
    \\

\citet{Guo2025CodeMirage}
    & Generated and paraphrased code
    & Known generated/paraphrased provenance
    & Detection performance (F1, TPR at 10\% and 1\% FPR); robustness across languages, generator models, paraphrasing, and cross-model settings
    \\

\citet{chon2024functional}
    & Generated solutions to benchmark tasks
    & Known generation process
    & Functional correctness (Pass@k); pairwise token-, embedding-, and LLM-based code similarity; Sim@k and CSim@k; similarity-weighted correctness (DPass@k)
    \\

\citet{Cynthia2026BeyondBugFixes}
    & Agent bug-fix PRs: base versus merge
    & Identified agent PRs
    & Newly introduced SonarQube issues: frequency/count, churn-normalized density, severity, category, and rule-level prevalence
    \\

\citet{Yu2023}
    & Evolving Java projects
    & None (non-AI setting)
    & SonarQube violation occurrence and closure; fix rate and fix time by rule/type, category, and severity; validation of closed-versus-fixed violations
    \\

\citet{Tufano2026DevelopersAG}
    & Repeated study of repository disclosures
    & Self-admitted usage
    & Self-admitted ChatGPT/Copilot use in commits, issues, and PRs; usage-task categories and frequencies; reported usefulness of the assistance
    \\

\citet{Robbes2026AgenticMA}
    & GitHub repository traces
    & Heuristic agent-activity signals \citep{robbes2026promises}
    & Adoption prevalence and timing; file-level adoption and agent-assisted commit ratio; adoption across project contexts; commit size (files/lines changed) and commit-purpose categories
    \\

\midrule
This study
    & Same mature projects, 2021/24/26
    & Cohort evidence; no function labels
    & Semantic similarity (CodeBERT and CodeT5 cosine similarity); syntactic similarity (CodeBLEU*); maintainability (Code Smells, Duplicated Lines, SQALE Index, Comment Lines); complexity (Cyclomatic and Cognitive Complexity); reliability (Bugs); security (Vulnerabilities and Security Hotspots); structural growth (annualized growth of NCLOC, functions, classes, files, and statements); development activity (commits, lines added, and lines removed)
    \\
\bottomrule
\end{tabularx}
\end{table*}
\paragraph{Longitudinal studies of AI assistance.}
Longitudinal repository analysis is already an active line of research. \citet{Xu2026CodeTransformed} examine style, quality-related properties, and similarity on a large-scale dataset of scientific open-course code. They report changes in naming conventions, complexity, maintainability, and similarity that align with characteristics of LLM-generated code.
\citet{He2026SpeedCostQuality} and \citet{Agarwal2026AIIDEsAutonomousAgents} study the impact of AI-enabled IDEs, such as Cursor AI, on the productivity of developers with a controlled user study \citep{He2026SpeedCostQuality} and a longitudinal study over open-source repositories \citep{Agarwal2026AIIDEsAutonomousAgents}. 
They measure monthly repository outcomes after AI adoption, including development velocity, static-analysis warnings, cognitive complexity, and comment density. 
\citet{Fang2026Conversation} study open-source coding conversations and repository histories, examining monthly indicators around adoption. They find that bug-fix, test-related, and CI-related code do not show deterioration after AI adoption. These results correlate with our findings for RQ2.

\paragraph{Code quality and similarity.}

\citet{Shi2025} characterize lexical similarity, conciseness, formatting, and naturalness in machine- and human-authored code. Such known-origin comparisons allow only for establishing measurable differences under the studied conditions, but are difficult to generalize to actual open source software. 
In this sense, \citet{Guo2025CodeMirage} evaluate various AI code detectors under generated and paraphrased code. They find mixed results, motivating caution about code provenance attribution.

Regarding the study of generated code similarity, \citet{chon2024functional} examined diversity across generated solutions in addition to functional correctness for a set of known programming tasks. They highlight that the models of that time (GPT-3.5-turbo and equivalent) usually produced functionally correct code that was also similar to each other. This seems to be changing with more recent LLMs, mainly for Java, as shown in our RQ1 results.

Regarding code quality, \citet{Cynthia2026BeyondBugFixes} used SonarQube to compare base and merged revisions of agent-generated Python bug-fix pull requests. They examine newly introduced issues and show that merge success does not reliably reflect post-merge code quality, highlighting a need for better checks for agent-generated bug-fix pull requests. This is related to our findings for RQ2. 
Static-analysis measures have also been studied outside the generative-AI setting. Notably, \citet{Yu2023} trace SonarQube violations and examine their fixes across evolving Java projects. This supports the use of rule findings as an empirical object and thus corroborates our conclusions. 

\paragraph{Mining visible LLM and coding-agent activity.}

\citet{Tufano2026DevelopersAG} analyze self-admitted generative-AI use in commits, issues, and pull requests, extending an earlier study to characterize reported tasks and usage. They show how the usage avenues of AI tools in open source projects have expanded, creating a taxonomy of 64 different usage tasks. 
\citet{Robbes2026AgenticMA} study the adoption of coding agents through GitHub repository traces, based on heuristics presented in \citep{robbes2026promises}. They find an estimated adoption rate of around 25\% in February 2026, and span a large range of programming languages and project topics. They study structural differences such as commit sizes that are larger in commits assisted by coding agents.

\paragraph{Positioning.}

\tabref{tab:closest-related-work} contrasts the closest empirical studies with ours in terms of study material, evidence of AI use, and measured outcomes. Our study complements these works by jointly examining three dimensions of software evolution across pre-LLM, potentially LLM-assisted, and detected agentic development eras. These are function-level semantic and syntactic similarity~(RQ1), multiple dimensions of software quality~ (RQ2), and structural growth and development activity~(RQ3), as detailed in \secref{sec:method}. Unlike studies focused on generated code, individual agent contributions, or adoption-related events, we compare historical snapshots of the same long-lived open-source projects, allowing us to examine how these characteristics change within established codebases over time. We also distinguish temporal evidence of potential LLM assistance from repository-specific evidence of agentic activity, without assuming that all code produced during a given era was AI-generated.

Although our observational design does not allow us to attribute the observed changes directly to AI adoption, we investigate whether changes across development eras align with expectations and concerns raised by prior research on AI-assisted development. Specifically, we examine whether similarity increases, whether software quality deteriorates, and whether structural growth or development activity changes as development practices evolve. By considering these outcomes together and comparing trends across Java and Python, our study provides a broader empirical perspective on how the transition toward LLM-assisted and agentic development coincides with the evolution of long-lived open-source software systems.
\section{Conclusion and Future Work}
\label{sec:conclusion}

This study examined code similarity, software quality, structural growth, and development activity across pre-LLM, potentially LLM-assisted, and detected agentic development eras in 50 long-lived open-source projects implemented in Java and Python. Results show that similarity trends vary by language and representation, with significant decreases in selected Java comparisons. Software quality remained broadly stable, with reductions in code-smell and technical-debt density between the pre-LLM and agentic eras. No statistically significant changes were found in structural growth or development activity.

Future work should investigate how repository-level instructions, prompts, and agent configurations influence code similarity, reuse, and diversity. Another direction is to examine whether changes in static-analysis metrics translate into differences in correctness, test effectiveness, security, and long-term maintainability. Combining fine-grained development histories with pull requests, issue discussions, and explicit AI-contribution provenance could help distinguish generated functionality from refactoring and maintenance, and clarify how developers review and integrate AI-generated code. The impact of version control on agent-driven development also needs to be investigated, including whether technologies such as Git restrict or limit the development process. Finally, expanding the analysis to more projects and programming languages and linking development activity to delivered functionality and maintenance effort could provide a more comprehensive understanding of the relationship between AI-assisted development and software evolution.

\section*{Declarations}

\subsection*{Funding}
This work was partially funded by the Natural Sciences and Engineering Research Council of Canada (NSERC), grant number RGPIN-2025-05677.

\subsection*{Ethical approval}
Not applicable, as this research did not involve human participants or interventions and did not raise any ethical concerns.

\subsection*{Informed consent}
Not applicable, as this research did not involve human participants.

\subsection*{Author Contributions}
Luciano Marchezan, Romain Lefeuvre, and Kevin Delcourt contributed to the conceptualization, methodology, investigation, data analysis, and writing of the manuscript. Benoit Combemale, and Houari Sahraoui contributed to the research through their respective roles in the study design, methodology, interpretation of results, supervision, and manuscript review and editing. All authors read and approved the final manuscript.

\subsection*{Data Availability Statement}
A replication package containing the data and analysis artifacts for this study is available in an online repository~\citep{EMSEGenAIReplication}. The source code is also available on Software Heritage~\citep{reproductionPackageScript}.

\subsection*{Conflict of Interest}
Considering the EMSE journal editorial board, possible conflicts of interest include: Bram Adams, Foutse Khomh, Ali Ouni, Jinqiu Yang, and Paul Gruenbacher.

\subsection*{Clinical trial number}
Not applicable.

\bibliographystyle{spbasic}   
\bibliography{bib/references.bib}

\end{document}